# Scaling from traits to ecosystems: Developing a general Trait Driver Theory via integrating trait-based and metabolic scaling theories

**Short title:**   Scaling from traits to ecosystems

**Authors:**   Brian J. Enquist*   (benquist@email.arizona.edu)[1,2]

Jon Norberg   (jon.norberg@ecology.su.se)[3,4]

Stephen P. Bonser   (s.bonser@unsw.edu.au)[5]

Cyrille Violle   (cyrille.violle@cefe.cnrs.fr)[1,6]

Colleen T. Webb   (Colleen.Webb@ColoState.edu)[7]

Amanda Henderson   (amandahenderson@email.arizona.edu)[1]

Lindsey L. Sloat   (llsloat@email.arizona.edu)[1]

Van M. Savage*   (vsavage@ucla.edu)[2,8,9]

[1]Dept. of Ecology and Evolutionary Biology, University of Arizona, Bioscience West, Tucson, AZ 85721, USA.

[2]Santa Fe Institute, 1399 Hyde Park Rd, Santa Fe, NM 87501, USA

[3] Dept. of Systems Ecology, Stockholm University, 18697 Stockholm, Sweden

[4] Stockholm Resilience Centre, Stockholm University, 18697 Stockholm, Sweden

[5] Evolution and Ecology Research Centre and School of Biological, Earth and Environmental Sciences, University of New South Wales, Sydney NSW, 2052, Australia

[6]CNRS, Centre d'Ecologie Fonctionnelle et Evolutive, UMR 5175, Montpellier, France

[7]Dept. of Biology, Colorado State University, Fort Collins, CO 80524, USA

[8]Dept. of Biomathematics, David Geffen School of Medicine at UCLA, Los Angeles, CA 90095

[9]Dept. of Ecology and Evolutionary Biology, UCLA, Los Angeles, CA 90095

\* Contributed equally to the paper

| | |
|---|---|
| **Corresponding author:** | Brian J. Enquist |
| **Affiliation:** | Department of Ecology and Evolutionary Biology |
| | University of Arizona |
| **Address:** | 1041 E. Lowell St., Tucson, AZ 85721 |
| **Phone:** | (520) 626-3336 |
| **Email:** | benquist@email.arizona.edu |

**Abstract**

**Aim:** More powerful tests of biodiversity theories need to move beyond species richness and explicitly focus on mechanisms generating diversity via trait composition and diversity. The rise of trait-based ecology has led to an increased focus on the distribution and dynamics of traits in communities. However, a general theory of trait-based ecology, that can apply across different scales (e.g., species that differ in size) and gradients (e.g., temperature), has yet to be formulated. While research focused on metabolic and allometric scaling theory provides the basis for such a theory it does not explicitly account for differences traits within and across taxa, such as variation in the optimal temperature for growth. Here we synthesize trait-based and metabolic scaling approaches into a framework that we term "Trait Drivers Theory" or TDT. It shows that the shape and dynamics of trait distributions can be uniquely linked to fundamental drivers of community assembly and how the community will respond to future drivers. To assess predictions and assumptions of TDT, we review several theoretical studies, recent empirical studies spanning local and biogeographic gradients. Further, we analyze how the shift in trait distributions influences ecosystem productivity across an elevational gradient and a 140-year long ecological experiment. We argue that our general TDT provides a baseline for (*i*) recasting the predictions of ecological theories based on species richness in terms of the shape of trait distributions; and (*ii*) integrating how specific traits, including body size, and functional diversity 'scale up' to influence the dynamics of species assemblages across climatic gradients and how shifts in functional composition influences ecosystem functioning. Further, it offers a novel framework to integrate trait, metabolic/allometric, and species-richness based approaches in order to build a more predictive functional biogeography to show how assemblages of species have and will respond to climate change.

## I. Introduction

Understanding and explaining species richness patterns have had far-reaching influence on the development of ecology. Biodiversity science strives to understand the drivers and consequences of variation in the number of species, and how species abundances change across spatial and temporal scales (MacArthur, 1972; Rosenzweig, 1995). These changes in species richness have also been linked with changes in ecosystem functioning. The Biodiversity Ecosystem Functioning (BEF) hypothesis states that ecosystems with greater biodiversity are more productive and stable (Naeem *et al.*, 1994; Tilman *et al.*, 1997; Tilman, 2001). Attempts to answer these questions have led to debates that polarized the field (Wardle, 2002), and a growing consensus that species numbers alone do not inform us about all important aspects of ecosystem functioning and community responses to environmental change (Chapin *et al.*, 2000; Díaz & Cabido, 2001; Stevens *et al.*, 2003; Diaz *et al.*, 2007).

More recently, trait-based approaches have focused on recasting classical questions from the species richness literature (Lavorel & Garnier, 2002; McGill *et al.*, 2006; Violle *et al.*, 2007; Hillebrand & Matthiessen, 2009; Lamanna *et al.*, 2014). Instead of species richness, there is an attempt to focus on functional traits and diversity in trait values (Díaz & Cabido, 2001; Lavorel & Garnier, 2002; Petchey & Gaston, 2002; Mason *et al.*, 2005; Roscher *et al.*, 2012). In addition, metabolic scaling theory or MST has focused on the central role of body size as a critical driver of ecological, ecosystem, and evolutionary patterns and processes (Enquist *et al.*, 1998; Enquist *et al.*, 2003; Savage *et al.*, 2004; Gillooly *et al.*, 2005). One could also ask about diversity in the number and/or range of trait or body size values and to some degree, this depends on how traits are defined. The premise is that measures of traits, including body size, can better reveal the mechanisms and forces that ultimately structure biological diversity (Grime, 2006; McGill *et al.*, 2006; Stegen *et al.*, 2009) and increase the generality *and* predictability of ecological models (Díaz *et al.*, 2004; Webb *et al.*, 2010; Kattge *et al.*, 2011). Trait-based approaches have especially received attention for plant life histories and strategies due to a renewed interest in measuring traits across different environments and scales (Craine, 2009). While this has long been part of comparative physiology and ecology (see Grime, 1977; Arnold, 1983), it is now being heralded as its central paradigm (Westoby & Wright, 2006; Craine, 2009). Similarly, trait-based approaches are being used to disentangle the forces that structure larger scale biodiversity gradients (Reich & Oleksyn, 2004; Han *et al.*, 2005; Reich, 2005; Swenson & Enquist, 2007; Safi *et al.*, 2011; Belmaker & Jetz, 2013) and to predict large-scale ecosystem shifts due to climate change (Elser *et al.*, 2010; Frenne *et al.*, 2013).

*Central limitations of trait-based ecology*

An important limitation to developing a more predictive trait-based ecology is that its focus and implementation has relied almost entirely on empirical correlations and null models (for example see discussion in Swenson, 2013). There is a need for theory and quantitative arguments to move beyond pattern searching. Further, trait-based ecology has largely developed independently from metabolic scaling theory where the role of body size – arguably a key trait - is central to scaling up organismal processes. Nonetheless, a key focus of trait-based ecology is to identify the general processes underlying trait-based ecology (Suding *et al.*, 2008b; Enquist, 2010; Shipley, 2010; Webb *et al.*, 2010; Weiher *et al.*, 2011). Such an advance would help guide the explosion of trait-based data collection (Kattge *et al.*, 2011; Dell *et al.*, 2013), develop a more predictive ecology, and to organize rapidly developing directions in trait-based ecology (McGill *et al.*, 2006; Funk *et al.*, 2008; Suding *et al.*, 2008b; Shipley, 2010; Lavorel *et al.*, 2011; Boulangeat *et al.*, 2012).

Another limitation is the debate about whether biodiversity, trait diversity, or both are important for ecosystem functioning (Loreau *et al.*, 2001; Hooper *et al.*, 2004). We agree with Cardinale et al. (2007) that this debate is largely a false dichotomy. Increasingly, the evidence shows that both the number of species and types of species in an ecosystem impact biomass production For example, focusing solely on species number, however, has resulted in sometimes positive, negative or null relationships between species richness and ecosystem functioning (Grace *et al.*, 2007; Roscher *et al.*, 2012).

Lastly, because trait-based ecology measures properties of individuals that are linked to the environment and because it attempts to make predictions for ecosystem functioning, it must be able to scale from individuals to ecosystems. However, achieving this requires an exciting but extremely challenging synthesis of physiology, population biology, evolutionary biology, community ecology, ecosystem ecology, and global ecology (Webb *et al.*, 2010; Reich, 2014). In this paper we suggest combining trait-based approaches with metabolic scaling theory to make some progress on this problem.

Here, we present a novel theoretical framework to scale from traits to communities to ecosystems and to link measures of diversity. We argue that trait-based ecology can be made more predictive by synthesizing several key areas of research and to focus on the shape and dynamics of trait distributions. Our approach is to develop more of a predictive theory for how environmental changes, including land use and shifts in abiotic factors across geographic and temporal gradients, influence biodiversity and ecosystem functioning (Naeem *et al.*, 2009). We show how starting with a few simple but general assumptions allows us to build a foundation by which more detailed and complex aspects of ecology and evolution can be added. We show how our approach can reformulate and generalize the arguments Chapin et al. (2000), McGill et al. (2006), and Violle et al. (2014) by integrating several insights from trait-based ecology (Garnier & Navas, 2012) and metabolic scaling theory (West *et al.*, 1997; Enquist *et al.*, 1998; Gillooly *et al.*, 2001). In doing so, we can derive a synthetic theory to begin to (i) assess differing assumptions underlying the assembly of species; (ii) assess the relative importance of hypothesized drivers of trait composition and diversity; and (iii) build a more predictive and dynamical framework for scaling from traits to communities and ecosystems. We call this theory, Trait Driver Theory or TDT, because it links how the dynamics of biotic and abiotic environment then drive the performance of individuals and ecosystems via their traits. Combining MST with trait-driver approaches allows TDT to work across scales also addresses one of MST's key criticisms: it does not incorporate ecological variation – such as trait variation - and cannot be applied to understanding the forces that shape the diversity and dynamics of local communities (Tilman *et al.*, 2004; Coomes, 2006).

**II. Trait Driver Theory**

TDT is based on a synthesis of three influential bodies of work. The first are trait-based approaches that are largely encapsulated in Grime's Mass Ratio Hypothesis or MRH (Grime, 1998). The MRH states that ecosystem functioning is determined by the characteristics or traits of the dominant (largest biomass) species. Implicit in the MRH is the idea that traits of the dominant species are a more relevant measure than species richness. The second component is the generalized and quantitative approach to trait-based ecology through Norberg et al. (2001) who used a mathematical framework to link the distribution dynamics of phenotypic traits with environmental change and ecosystem functioning (Norberg *et al.*, 2001; Norberg, 2004; Savage *et al.*, 2007; Shipley, 2010). The third component is Metabolic Scaling Theory (MST). MST can be used to predict how variation in organismic size and the traits associated with metabolism will then influence individual performance (growth and resource use) and how these performance measures will then scale up to influence communities and ecosystems (Enquist *et al.*, 1998; Enquist *et al.*, 2003; Enquist *et al.*, 2009; Yvon-Durocher *et al.*, 2012). MST achieves this by showing how variation in individual rates of mass growth, $dM/dt$, and metabolism, $B$, can linked to variation in a few key traits (e.g. body size, $M$, and traits related to cellular metabolism and allocation; see West *et al.*, 2002; Enquist *et al.*, 2007b; Enquist *et al.*, 2009; von Allmen *et al.*, 2012).

*The central assumptions of trait based ecology and the 'Holy Grail' of trait-based ecology* - Trait-based ecology assumes that there are traits that are functional, meaning they link the environment to variation in whole-organism performance and ultimately fitness (Violle *et al.*, 2007; see Fig.1). That is, as shown in Fig. 1, variation in traits influence organismal performance (e.g., metabolism, growth rate, demographic rates, etc.) and ultimately fitness (Ackerly & Monson, 2003; Garnier *et al.*, 2004; Lavorel *et al.*, 2007; Violle *et al.*, 2007). This approach has been recently validated with a comparative study linking variation in individual-level traits with variation in life history and demography parameters (Adler *et al.*, 2014). Another key assumption of trait-based ecology is that traits of individuals can be used to predict individual performance that can be effectively summed or scaled up to the functioning of ecosystems (Lavorel & Garnier, 2002; Suding *et al.*, 2008b). The *raison d'être* and the 'Holy Grail' of trait-based ecology is to use functional traits, rather than species identities, to better predict community and ecosystem dynamics (Lavorel & Garnier, 2002; Lavorel *et al.*, 2007; Suding & Goldstein, 2008).

*Linking traits, individual performance, communities, and ecosystem functioning*
We start by extending Grime's MRH. Grime argued that dominant traits rather than species number drive ecosystem functioning, in order to identify the dominant traits. As a result, it is crucial to measure the trait frequency distribution defined by biomass for the assemblage. An important question is in order to assess the MRH should one use abundance or biomass weighted mean trait values to best estimate the frequency distributions. TDT also focuses on the trait frequency distribution $C(z)$ — the histogram of biomass across individuals characterized by a given trait value, *z*, summed across all individuals within and across species. Thus, *C(z)* captures both intra- and interspecific trait differences. However, unlike the MRH use of the mean and current interest in using community-weighted mean traits, we are interested in the overall shape of the distribution of phenotypes described by the moments—variance, skewness, kurtosis—beyond the mean. We can link individual growth rate and the population per capita growth rates via how traits influence organismal performance via the growth function,

$$f(z) = \left[\frac{1}{C(z)}\right]\left[\frac{dC(z)}{dt}\right]$$

(1)

where $C(x)/dt$ is the biomass growth rate for *all* individuals with a given trait value *z* (see Supplemental Document). By integrating the growth equation across all values of the trait across individuals, we can derive dynamic equations for how total community biomass, $C_{Tot}$, depends on the shape of the biomass-trait distribution, *C(z)*, and how that shape itself changes in time. Consequently, the Net Primary biomass Production or NPP is

$$\frac{dC_{Tot}}{dt} = \int f(z)C(z)dz$$

(2)

(see also Norberg *et al.*, 2001; Lavorel & Garnier, 2002; Vile *et al.*, 2006). Eqn. 2 requires understanding what sets the form of *f(z)*.

*Linking dynamics of trait distributions to environmental change and immigration -*

Starting with Norberg et al. (2001) and Savage et al. (2007), we focus on how traits that strongly influence organismal growth rate are influenced by the environment, $E$. In Fig. 2 we show an example of how variation in a given trait, $z$, translates to variation in per capita growth rate across an environmental gradient. This example assumes that all individuals in the assemblage ultimately compete for similar limiting resources, and that, there is an optimal environment where growth is fastest (Fig. 2). Although our approach starts with a single trait, trait-based models can straightforwardly incorporate multiple, correlated traits (Savage *et al.*, 2007 ; see also Supplemental Document). By incorporating temporal environmental forcing into the growth function, TDT predicts how the distribution of traits, $C(z)$, responds to both biotic and abiotic drivers (Norberg *et al.*, 2001; Savage *et al.*, 2007).

The shape and the dynamics of the trait distribution ultimately reflect a balance between two rates – the introduction/immigration of traits, $I$, into an assemblage and the outcome of variation in the performance, $f$, of those traits within the assemblage. A general trait-based equation for growth and immigration is given by

$$\frac{dC(z)}{dt} = f[z, E, C(z)]C(z) + I[z, E, C(z)]$$

(3)

Here, we now explicitly include the effects of the environment, $E$, and the trait distribution $C(z)$ as part of the growth, $f$, and immigration, $I$, functions because they can influence both via environmental change, competition, facilitation, sampling effects or other biological interactions such as density dependence (Savage *et al.*, 2007). The second term, the immigration function, *I*, reflects the external input of individuals into the community stemming from dispersal as well as the introduction of traits into the assemblage from other factors including evolutionary processes (mutations) and potential seed banks.

Next, we use two assumptions to constrain the form of eqns 2-3. First, a central tenet of TDT and a well-grounded concept in ecology and evolution is that across an environmental gradient, *E*, organisms will tend to have a unimodal functional response in their performance and fitness functions (Fig 2). As a result, a shift in the environment, *E*, will affect the per capita population growth rate and thus which traits are dominant in the community or assemblage (Whittaker *et al.*, 1973; Davis & Shaw, 2001) and the rate of trait evolution (Levins, 1968). Second, there are specific traits that link environmental drivers to individual growth rate, and the trait driven per capita biomass growth rate, *dC(z)/dt* (Fig. 1; (Arnold, 1983)). The performance or growth function *f(z)* then is a result from an environment mediated tradeoff between traits, such as investment in growth versus defense or from investment in growth rate versus desiccation resistance. As a result, for a constant environment, *E*, there are optimal trait values, $z_{opt}$, that maximize the growth function given the environment.

In the case of a single trait optimum, we approximate this as a symmetric function such as a Gaussian or quadratic tradeoff $f(z) \propto \left[1 - \left(\frac{z-z_{opt}}{\sigma^2}\right)^2\right]$ where $\sigma^2$ is the trait breadth of the tradeoff function. If the environment is constant and immigration, *I*, is zero, individuals with traits that match, $z_{opt}$, will gradually replace all other individuals, and the trait distribution will collapse on a single point for the optimal trait value $z_{opt}$ (Norberg et al. 2001). Thus, TDT is consistent with a competitive trait hierarchy view of assemblage interactions (Goldberg & Landa, 1991; Freckleton & Watkinson, 2001; Mayfield & Levine, 2010; Kunstler *et al.*, 2011) as well as a population source-sink view of assemblage (Pulliam, 1988) and metapopulation perspective of trait dynamics across environmental gradients (Davis & Shaw, 2001). As we discuss below, additional biotic and abiotic interactions and processes can also be shown to influence the shape of the trait distribution via growth function and immigration (Weiher & Keddy, 1995).

**III. Predictions of TDT –**

Next, we emphasize the central predictions of Trait Driver Theory. These predictions are also summarized in Table 1 in terms of how different measures of the trait distribution can provide novel insight and predictions regarding the main drivers of the current composition of the species assemblage as well as the future dynamics of the species assemblage.

***Prediction (1)****: Shifts in the environment will cause shifts in the trait distribution (Fig. 3; Table 1).*

***Prediction (2)****: The difference between the optimal trait and the observed mean trait, as well as the trait variance, provide a measure of the capacity of a community to respond to environmental change (Fig 3 and S1; Table 1).*

Shifts in the abiotic or biotic environments, represented by *E* and *C(z)* respectively, will lead to corresponding shifts in the community trait distribution. The magnitude of the shift over some time and the rate of change of the shift can both be calculated from eqn 2. According to eqn 2, the value of the optimal trait will change with the environment (e.g. Ackerly, 2003), while the mean trait of the community will approach the optima but with a lag in time according to how long it takes for either trait plasticity and/or the processes of species sorting and selection to act (Ghalambor *et al.*, 2007). In environments where the optimal trait value is changing quickly relative to generation times or plasticity, there may be little capacity for the mean community trait, $\bar{z}$, to track these changes. In such circumstances, $\bar{z}$, may never, or only rarely, be expressed at an optimal value for the current environment. Nonetheless, we expect that for most communities, the difference between the optimal trait, $z_{opt}(E)$, and the observed mean trait, $\bar{z}$, or $\Delta(E)$, will be a measure of how the community has responded/will respond to environmental change. Norberg et al. (2001) derive the general expression

$$\Delta(E) = z_{opt}(E) - \bar{z}$$

(4)

where $\Delta(E)$ quantifies the community trait "lag" in relation to the current environment. This measure is analogous to the 'lag load' in evolutionary theory (Maynard Smith, 1976). We can thus define $d\Delta/dt$ as the response capacity of a community. Eqn. 4 predicts that capability of the assemblage to respond to directional shifts in the environment will be directly proportional to the trait variance, $d\Delta/dt \propto V$. Importantly, within TDT, directional selection for more optimal trait values need not always lead to an *increase* in per capita growth rate, *f*. Because of trade-offs between traits and frequency and density-dependent effects on performance and fitness, the performance (and fitness) associated with the new optimum value likely differs from the fitness and growth rate in the previous environment (Antonovics, 1976; Dieckmann & Ferrière, 2004; Ferriere & Legendre, 2013). For example in Fig. 3 we highlight a hypothetical example of a shift in the community trait distribution from wet to dry that comes with a decrease in optimal performance. Extensions of TDT can in principle include these effects (Savage *et al.*, 2007).

In sum, predictions 1-2 formalize Chapin *et al.* 's conceptual framework (Chapin *et al.*, 2000) for the development of a predictive trait-based ecology. In the case of multiple traits underlying growth, *f*, differing trait combinations could lead to similar growth rates in differing environments (see also Marks & Lechowicz, 2006). We note that these predictions implicitly ignore the effects of frequency dependence but elaborations of TDT can include these effects (see eqn. 5 in Savage et al. 2007).

***Prediction (3)****: The skewness of the trait distribution can be an indicator of past or ongoing immigration and/or environmental change due to lags between growth, reproduction, and mortality* (Table 1).

Because of time lags between environmental change and the time scale of organismal responses (growth, demography etc.), the trait distribution of an assemblage will not be able to instantaneously track environmental change, and skewness in the trait distribution will develop (see also Figs. 3, S1). Alternatively, skewness may reflect differential immigration of traits from one side of a habitat or a community that contains 'sink' populations (Pulliam, 1988) supported via immigration. Neutral theory (Hubbell, 2001), in which traits have no demographic effects, could also lead to skewed distributions due to neutral trait evolution. The implication is that trait-based ecology can infer the dynamics of trait assemblages via assessing the shape of contemporary trait distributions. Combining information on the shape trait distributions with additional information such as dispersal history and/or size distributions would help separate lag effects from drift and differential immigration.

***Prediction (4)***: *The rate of change of net ecosystem productivity in response to environmental change can be predicted via the growth function, f, and the shape of the community biomass-trait distribution C(z) at some initial time* (Table 1)

In the simple case of a single trait with a single environmental driver, Norberg et al. (2001) derived a general expression linking the dynamics of the trait distribution by noting eqn. 2, equation 3 can be approximated as

$$\frac{dC_{Tot}}{dt} \approx \left[ f(z,E,C(z))_{z=\bar{z}} + \frac{\partial^2 f(z,E,C(z))}{\partial z^2}\bigg|_{z=\bar{z}} V \right] C_{Tot} + I$$

(5)

Eqn. 5 follows from a Taylor expansion that effectively linearizes the equations. If the terms in brackets depend on total biomass, $dC_{Tot}/dt$ would scale non-linearly with total biomass, but in the simplest case, these terms are independent of total biomass implying that production scales linearly with total biomass. In eqn. 5 the net primary production, $dC_{Tot}/dt$, is equal to the growth rate of the mean community trait, $\bar{z}$, plus the second term that accounts for how much variation there is in the community trait distribution, *V*. Because the growth function, *f(z,E)*, has a maximum at $z_{opt}$, we expect the second derivative term to be negative, as long as $\bar{z}$ is in the neighborhood of $z_{opt}$ (Norberg et al. 2001; see also discussion in Supplemental Document) reflecting the increasing reduction in growth rate as trait values increasingly differ from $z_{opt}$, (see Fig. 2 and eqn. 4). The unimodal shape is the simplest assumption requiring only the mean and variance. There is reason to expect that *f* can be approximated as unimodal. For example, growth rates typically exhibit unimodal response with measures of temperature, pH, etc. (McGill *et al.*, 2006). Again, the term *I* gives the addition of biomass through immigration/dispersal.

***Prediction (5)****: Within a community whose growth rate depends on a single trait, an increase in the variance of that trait will lead to a decrease in net primary production (Table 1).*

An additional prediction is that for communities whose growth is driven by a single key trait, larger trait variance, *V*, will decrease the net primary production for the whole community because a higher proportion of individuals differ from $z_{opt}$ (Norberg *et al.*, 2001). This idea of a tradeoff between short term productivity and long-term response to environment is reflected by agricultural imperatives with agricultural issues where short-term productivity is emphasized and variance in traits is minimized in trait values and short term productivity is maximized. Elaborations of TDT have shown that incorporating multiple limiting resources, multiple traits and trait covariation (Savage et al. 2007) can weaken, nullify, or even reverse the predicted negative relationship between $dC_{Tot}/dt$ and $V$.

From eqns 3 and 4, the rate at which a community can track environmental change will be greater when there is greater trait variance. Intuitively, greater variance leads to more extreme traits being immediately available to respond to environmental change. Thus, the rate of response will also depend upon the specific form of the growth function $f$ (e.g., for a given value of $z$, how does $f$ vary across an environmental gradient?; (see Savage *et al.*, 2007 and discussion in Supplemental Document). For example, if $f$ is a simple Gaussian or polynomial function with $E$ (Fig 1), then the value of $d\Delta/dt$ can be approximately proportional to the community trait variance, $V$. Building on the work of Norberg et al. (2001) and Savage et al. (2007), these equations can be extended to include higher-order moments such as skewness and kurtosis.

*Extending TDT via recasting and assessing different ecological hypotheses about diversity*

As TDT predicts that over time only one phenotype should dominate a given area characterized by a given environment and important question is what maintains diversity (trait variation) within an assemblage? According to TDT trait variance can be increased by many differ ways. Immigration, $I$, from outside the assemblage, as well as from a directionally shifting or a temporally variable environment (Norberg et al. 2001; Savage et al. 2007) can increase and/or maintain trait variation. Further, theoretical elaborations of TDT have shown that the diversity of phenotypes (traits) present in a given assemblage can be influenced by tradeoffs between traits that influence growth. For example, tradeoffs between allocation to predator defense and growth rate (Norberg et al. 2001; Savage et al. 2007) can increase the variance of a given trait. In a variable environment, correlations between traits that underlie the growth function, $f$, leads to the survival of organisms with trait values that are less favorable in the current environment but may be well suited for new environments that arise. Thus, phenotypic trait correlations among traits can ramify to have quantitative effects on ecosystem dynamics (lowering NPP) and enable assemblages to better track environmental change (Savage et al. 2007).

Additionally, trait variation can also stem from additional ecological hypotheses for biological diversity. An exciting aspect of TDT is that differing ecological hypotheses based on species richness can be recast in terms of traits. In Table 1 and Fig. S1 we overview the predictions of the different theories as recast in the light of TDT. As a result, TDT can then be used to 'scale up' the implications of many differing classic and current hypotheses for species richness via the assumptions of trait distributions implicit in these theories. Thus, TDT also offers a starting basis to assess differing hypotheses regarding diversity, community dynamics and ecosystem functioning. Differing ecological theories based on species richness (neutral theory, abiotic filtering, competitive exclusion, Chesson's storage effect, or rare species advantages; see discussion Supplemental Document) as well as the effects of abiotic processes (shifts due to environmental change, disturbance) will uniquely influence the shape of the community trait distribution and the potential of the assemblage to maintain diversity, as reflected in changes in the trait variance, $dV/dt$. The relative strengths of abiotic, biotic, and neutral processes will lead to different shapes of trait distributions that will have different implications for responses of community to directional shifts in the environment as well as ecosystem functioning.

**IV. Extending and Parameterizing Trait Driver Theory**

*Scaling from individuals to ecosystems using Metabolic Scaling Theory*

So far, Trait Driver Theory (TDT) assumes that there is no variation in organismal size. Instead, the total biomass associated with a trait, $z$, is denoted by $C(z)$. This notation avoids ever needing to account for *individual organismal* mass, $M$, or even the number of individuals with mass. However, body size can vary greatly – it is also an important trait that influences variation in organismal metabolism (Peters, 1983), population growth rate (Savage et al 2004), and abundance (Damuth, 1981; Enquist *et al.*, 1998). The scaling equations in Metabolic Scaling Theory or MST differ from TDT so far as they are phrased in terms of individual mass. In order to integrate these theories, we use three insights from MST (West *et al.*, 1997; Savage *et al.*, 2004; Enquist *et al.*, 2007b; Enquist *et al.*, 2009; West *et al.*, 2009), to explicitly formulate TDT to work across scales in organismal size, $M$, and environmental changes or gradients in temperature, $T$. As we show, MST provides the basis to formally link traits, organismal growth rate, and ecosystem fluxes (Enquist *et al.*, 2007a; Enquist *et al.*, 2007b).

First, MST is explicit about how the organismal growth is dependent upon the size of the organism. In the case of MST, we start with how organismal biomass growth rate, $dM/dt$, is related to whole-organism metabolic rate, $B$, and organismal mass, $M$, as

$$\frac{dM}{dt} = b_0(z)M^\theta$$

(6)

where $b_0(z)$ is a metabolic coefficient that depends on a single or set (meaning $z$ is a vector) set of traits. The allometric scaling exponent $\theta$ is hypothesized to reflect the branching geometry of vascular networks (Enquist *et al.*, 2007b). Theory and empirical data point to $\theta \approx 3/4$ for large size ranges (Enquist *et al.*, 2007c; Savage *et al.*, 2008). Eqn 6 has recently been shown to be a good characterization of tree growth (Stephenson *et al.*, 2014) and is a specific case of a more generic growth function (West *et al.*, 2001; Moses *et al.*, 2008) that can be applied to both plants and animals. While we focus here on a specific plant growth model, we note that other trait-based models have recently been developed for animals and phytoplankton (Ricker, 1979; Muller *et al.*, 2001; West *et al.*, 2001; Litchman & Klausmeier, 2008) and they could also be used to parameterize TDT. Below, we elaborate eqn. 6 to explicitly include the traits for plants that underlie $b_0$ and how we can use this equation as the basis for a general trait-based growth function.

To integrate MST into TDT, we first recognize that, $C(z)$, the biomass associated with trait $z$ can be expressed as $C(z) = \int dM\, C(z, M) = \int dM\, N(z, M)M$, where $C(z,M)$ is the mass density of individuals with both trait value $z$ and individual mass $M$, while $N(z,M)$ is the number density of individuals that have both trait value $z$ and individual mass $M$. In this expression, we have integrated over all possible values of mass, $M$, so that have the total biomass of *all* individuals with trait $z$. Furthermore, note that integrating this over all traits, $z$, gives the total biomass, $C_{TOT} = \int dz\, C(z) = \int dz \int dM\, N(z,M)M$.

To integrate MST into TDT we solve for the conditions of steady state where $N(z,M)$ is not changing in time. It can be shown (see Sup. Doc) the equation for the scaling of NPP with the total biomass of the assemblage is,

$$\frac{dC_{TOT}}{dt} = \langle b_0(z)M^{-\frac{1}{4}}\rangle_C\, C_{TOT}$$

(7)

where $dC_{Tot}/dt$ scales isometrically with $C_{Tot}$ and the $C$ subscript denotes that the average, denoted by brackets $<>$ is taken with respect to the biomass. This equation is in the most generic form of a general TDT equation. Note, that the TDT growth function, $f$, is now

$$f(z) = b_0(z)M^{-\frac{1}{4}}$$

(8)

and can be expanded and expressed in terms of the biomass-weighted central moments of the trait $z$, such as the variance, skewness, and kurtosis (see below). Again, the exponent, -1/4 is the idealized case and empirical values that may deviate from $\theta \approx 3/4$ can be used.

*Incorporating environmental tradeoffs in the growth function, f*

Eqn. (8) alone would predict that the per capita growth rate will increase forever as the trait, $b_0$, and mass-specific metabolic rate increase. In reality, though, there is some range of trait values at which the organism can grow. This is because there are tradeoffs in performance and fitness. Decreases in growth when, for a given value of $E$, the trait value gets either too small or too large. In the case of a given leaf trait such as leaf size or leaf investment (closely associated with variation in photosytnthetic rates and the specific leaf area), at some point, continued increases in leaf nitrogen may ultimately limit resource uptake as high $N$ would result in individuals more prone to herbivores, pathogens etc. and/or will result in water transport demands that would increasingly be maladaptive for a given local environment. Ultimately, one cannot have an infinitely large leaf, an infinitely thin leaf, or a plant that is all leaf area. Thus, deviation away from $b_{0,opt}$ would be associated with a tradeoff between specific trait values and plant performance (such as growth rate, survivorship and/or reproduction; (see also Ghalambor *et al.*, 2007)).

Incorporating tradeoffs between trait values, the environment, and performance is central to TDT. We can incorporate these tradeoffs in a general form by multiplying the scaling relationship by a quadratic function. As a result, $f(b_0)$ is maximal at the optimal trait, $b_{0,opt}$, and the niche width defined by $\sigma_{b_0}^2$ where

$$f(z) = b_0(z) M^{-\frac{1}{4}} \left(1 - \frac{(b_0 - b_{0,opt})^2}{\sigma_{b_0}^2}\right)$$

(9)

Here, the second term is the tradeoff function and $c_0$ is an overall constant coefficient. Expressing $f(b_0)$ across an environmental gradient, $f(b_0, E)$ would then reveal a unimodal growth function (Fig. 2).

The second insight from MST shows that the metabolic normalization, $b_0$, can be linked to specific traits. For the plant growth function, building on the insights from the relative growth rate literature (Evans, 1972; Lambers *et al.*, 1989; Poorter, 1989), Enquist et al. (2007b), derived an extension to eqn 6 that explicitly details the traits that together define $b_0$ and hence $f$ so that

$$b_0 \propto \frac{c}{\omega}\left(\frac{a_L}{m_L}\right)\dot{A}_L \beta_L$$

(10)

Eqn. 10 shows that, in addition to plant size, $M$, the rate of growth is governed by the scaling exponent, $\theta$, and five traits: (*i*) $\dot{A}_L$, the net leaf photosynthetic rate (grams of carbon per area per unit time); (*ii*) $a_L/m_L$, the specific leaf area or SLA, the quotient of area of the leaf, $a_L$, and the mass of a leaf, $m_L$; (*iii*) $\omega$, the carbon fraction of plant tissue; (*iv*) $c$ the carbon use efficiency of whole-plant metabolism, and (*v*) $\beta_L$, the leaf mass fraction (the ratio of total leaf mass to total plant mass) which is a measure of allocation to leaves. As a result, we can parameterize TDT with specific traits that underlie $b_0$

**V. Additional Predictions of Trait Driver Theory**

In the second column of Table 1 we summarize additional TDT predictions for scaling up community or assemblage trait distributions to predict several ecosystem level effects. Specifically, the shape of the trait distribution as measured via the central moments of the distribution

**Prediction (6)**: *Ecosystem net primary productivity, $dC_{Tot}/dt$ will scale with the total biomass but will be influenced differently by the mean and variance of the community trait distribution.*

A third insight from MST allows us to more formally link TDT with MST by including organismal mass dependence into TDT. In particular, most assemblages of organisms will be characterized by a distribution of sizes. For plants, following the arguments in Enquist et al. (2009), we can substitute the distribution of the number of individuals as a function of their size, $M$ or the size-spectra, $N(M)$. For the idealized case of $\theta \approx 3/4$, they show that $N(M) \propto M^{-11/8}$ and link the total biomass, $C_{Tot}$, with the size of the largest individual, $M_b$ where $M_b \propto C_{TOT}^{8/5}$. This allows us to consider a few special cases of Eqn. (7) that relate TDT and scaling equations already in the literature.

In the case of a given assemblage where there is no size distribution and only a single mass value, $M^*$, or a very small range of mass values, the scaling of NPP becomes

$$NPP = \frac{dC_{TOT}}{dt} = (M^*)^{-\frac{1}{4}} \langle b_0(z) \rangle_C C_{TOT}$$

(11)

The term $(M^*)^{-\frac{1}{4}}$ can be thought of as an overall normalization to the growth function *f(z)* from TDT. As such, this result reveals that TDT, as originally formulated (see Eqn 3), ignores variation in individual mass. Thus, based on Eqn. (6), growth functions within TDT should have a roughly $(M^*)^{-\frac{1}{4}}$ hidden with the normalization constant for their growth function. In the case where (*i*) organisms within the community or assemblage can differ greatly in their sizes; (*ii*) *z* and *M* are uncorrelated; and (*iii*) the number density is a separable function, such that $N(z,M) = N(M)N(z)$, it can be shown that the growth equation can be expressed two different ways. Each way depends on how one averages the trait distribution. In the first case we have,

$$\frac{dC_{TOT}}{dt} = k\langle b_0(z)\rangle C_{TOT}^{\frac{3}{5}}$$

(12)

and in the second case we have

$$\frac{dC_{TOT}}{dt} = k\langle b_0(z)\rangle \langle M^{-\frac{1}{4}}\rangle_C C_{TOT}$$

(13)

where *k* is a proportionality constant. Eqn 13 is equivalent to Eqn. (12) but expresses the growth function more in terms of the TDT framework such that the right side appears to have an overall *linear* dependence in $C_{TOT}$. As a result, in eqn. (12) we have a mixture of types of averages, with $\langle b_0(z)\rangle$ being the abundance average of the function *b₀(z)* while $\langle M^{-\frac{1}{4}}\rangle_C$ is the biomass average of $M^{-\frac{1}{4}}$. Both equations are equivalent ways to express the scaling of NPP function. Eqn. (12) is a more simple expression and only involves using the abundance average of the trait distribution. Eqn. 12 consolidates the organism mass average with the 3/5 scaling dependence of $C_{TOT}$. These derivations help to clarify when trait-based studies should use biomass or abundance weighted values in their studies.

Both equations assume a community steady state approximation where $N(z,M)$ is not changing in time. If this is violated (e.g. the community trait abundance or number distribution $N(z,M)$ is changing), then deviations from Eqns. 12-13 are expected. Nonetheless, these equations provide a basis for linking the scaling of organismal growth rate and trait variation of individuals with ecosystem-level processes. For all of these equations and cases, the functions inside the averages can be expended in terms of moments as done for TDT for biomass-weighted averages or as done in Savage (Savage, 2004) for abundance-weighted averages.

Putting all of this together with Eqn. (5) yields the prediction that equation 12 and 13 are then modified by the shape of the trait distribution, where for a given $E$, growth is reduced with departure from $b_{0,opt}$,

$$\frac{dC_{TOT}}{dt} \approx k \langle b_0(z) \rangle + \left[ \left(1 - \frac{(b_0 - b_{0,opt})^2}{\sigma_{b_0}^2}\right) \frac{d^2 f}{dz^2}\bigg|_{z=\bar{z}} V(b_0(z)) \right] C_{TOT}^{3/5}$$

(14)

The second term captures the reduction of production due to deviation from $b_{opt}$. Equation (14) represents a formal integration of foundations of TDT from Norberg et al. (2001) with MST from West et al. (1997) and Enquist et al. (2009). Because the growth function, $f(z,E)$, has a maximum at $z_{opt}$ (Fig 2) we expect the second derivative term, $\frac{d^2 f}{dz^2}$ to be negative, as long as $\bar{z}$ is in the neighborhood of $z_{opt}$. Importantly, Eqn.

14 enables one to parameterize TDT with a specific trait based growth function. Further, it enables the integration of physiological performance curves for how the key integrative trait, $b_0$, varies across a given environmental gradient, $E$. Eqn. (8) predicts that $dC_{Tot}/dt$ will increase with increasing community biomass, $C_{Tot}$. Note, here, the role of the relative breadth of species performance curves (see Fig. 1) is represented by $\sigma_{b_0}^2$.

***Prediction (7)****: Eqn. 14 and 16 generates specific and testable relationships for the scaling of trait means, dispersion, and ecosystem production (see Table 3).*

Eqn (14) predicts that there is a range of mean trait values for which increases in the variance of traits, $V(b_0)$, will *decrease* Net Primary Productivity, $dC_{Tot}/dt$, and shifts in $<b_0>$ will lead to corresponding shifts in $dC_{Tot}/dt$. This will occur whenever the mean trait value is near the maximum of the growth function, which should occur frequently because evolution is driving the mean trait to match the optimal trait with some lag time. However, there are also mean trait values for which increases in the variance of traits, $V(b_0)$, will *increase* Net Primary Productivity (NPP), $dC_{Tot}/dt$, and shifts in $<b_0>$ will lead to corresponding shifts in $dC_{Tot}/dt$. This will happen when the mean trait value is further from the optimal trait value and below an inflection point in the growth function that occurs for small trait values (see the example of a shift from historically wet climate regime to a dry regime in Figure 3). Intriguingly, this scenario suggests that trait variance can potentially act to either increase *or* decrease NPP depending on if the current trait distribution is close to the local trait optimum or not. So, if the assemblage is close to the optimal value, increases in trait variance will typically decrease NPP. This contrasts with biodiversity theories in which increasing variance (increased trait diversity) tends to increase NPP. Importantly, Eqn. 14 shows the influence of variation, *V,* of the traits that underlie $b_0$ observed within the community.

Integrating MST into a more generalized TDT lists several key traits for TDT. We explore predictions of TDT in the special case of a single trait driver such as specific leaf area, SLA=$(a_L/m_L)$. First, a change in the environment will likely be associated with a shift in the mean value of SLA. Second, using Eqn. 12 and 14 we expect that $\frac{dC_{Tot}}{dt} \propto (\langle SLA \rangle - \langle V_{SLA} \rangle) C_{TOT}^{3/5}$ . Thus, a shift that increases the abundance weighted mean trait value of <SLA> will lead[1] to an increase in NPP. Preliminary support of this prediction comes from empirical studies that have noted that increases in the mean community SLA is closely linked with increases in ecosystem productivity (Garnier *et al.*, 2004; Violle *et al.*, 2007) and follows from TDT. Third, due to the productivity-variance tradeoff predicted by TDT, an increase in the community variance in SLA or $V_{SLA}$, will lead to a *decrease* in productivity so long as the mean community SLA is close to the optimum value.

As we discuss below, TDT provides a foundation that can be modified by additional factors. When there are multiple trait shifts that may covary, measuring all of the traits listed in eqn. 11 would allow more detailed predictions. Of all the traits specified by eqn 11, there appears to be evidence that SLA may vary more across environmental gradients than other traits and be more important for linking changes in a trait driver or environment, *E,* with variation in local plant growth (see Supplemental Document). It varies across taxa (up to three orders of magnitude or approximately 1000 fold) and directionally varies across environmental gradients in soil moisture, irradiance, and temperature (see Garnier *et al.*, 2004; Wright *et al.*, 2005; Poorter *et al.*, 2009). SLA has also been noted to vary considerably *within* species in response to local changes in climate and abiotic conditions (Shipley, 2000; Cornwell & Ackerly, 2009; Jung *et al.*, 2010; Sides *et al.*, 2014).

---

[1] Note, because the growth equation *f* is for more instantaneous measures of growth this prediction is based on rates of more instantaneous NPP and not necessarily annual net primary production. So, accurate testing of this prediction with annual productivity data should make sure to standardize for growing season lengths.

**VI. Methods: Assessing TDT Assumptions and Predictions**

*Quantifying the shape of trait distributions* - In order to assess predictions of TDT, it is necessary to quantify the biomass distribution of traits, $C(z)$, in a species assemblage. This involves enough measurements of the trait values and body masses to obtain accurate estimates of the underlying distributions, as guided by sampling theory and statistics (Baraloto *et al.*, 2010; Paine *et al.*, 2011). The sampling must occur across all individuals within our group and thus incorporates both inter- and intraspecific trait variability (see Supplemental Document, and Violle *et al.*, 2012)). The sampling protocols often make choices that limit accurate measurements of within-species variability more than across-species variability. Indeed, simultaneous measurements of intra- and interspecific trait measures are rarely collected (Ackerly, 2003; Baraloto *et al.*, 2010). However, intra-specific variation in traits are important to determine the breadth of the distribution (Violle & Jiang, 2009; Sides *et al.*, 2014). Trait abundance or biomass distributions, $C(z)$, can be approximated through sampling (see Supplemental Document), so that predictions of TDT can be tested without explicitly measuring the traits of all individuals.

There are two reasonable approximations for community trait distributions. The first approximation method calculates the weighted trait distribution by taking the mean species trait value and multiplying by a measure of dominance (cover, biomass, abundance (Grime, 1998)). This method can be implemented by calculating the central moments of the joint-distribution. In the Supplemental Document we show the equations used to approximate the community trait moments in particular the community weighted mean, variance, skewness, and kurtosis respectively (CWM, CWV, CWS, and CWK). Community weighted metrics, however, ignore intraspecific variation. Increasingly it is becoming clear that intraspecific variation can contribute to a considerable amount of trait variation (Messier *et al.*, 2010) and that relying on species mean trait values may not provide a robust measure of the shape of a trait distributions (Violle *et al.*, 2012).

A second method utilizes sampling theories to help avoid the time-consuming work of sampling the traits of all individuals. Sub-sampling individuals can be used to better approximate how intraspecific variation influences the community distribution. In the Supplemental Document we develop this new method and provide code to implement this method (see discussion in SI). The method utilizes sub-sampling individuals to obtain a better approximation of how intraspecific variation influences the community distribution. By subsampling individuals to estimate intraspecific trait variation within each species, one can begin to incorporate intraspecific variation around mean trait values for each species. We expect that utilizing this method in addition to the incorporation of how MST influences the scaling of the number of individuals will improve estimates for the shape of trait distributions. In short, by subsampling individuals for each species within a given assemblage one can begin to incorporate intraspecific variation around mean trait values for each species.

*Testing predictions of TDT* - We tested several of the specific TDT predictions (Table 1) and assumptions using several examples that allow us to assess temporal and spatial variations of trait distributions. First, we searched the literature to determine if trait distributions measured from individuals actually do shift across local environmental gradients. Second, we assessed the dynamics of trait distributions and ecosystem carbon flux measures using data from an elevational gradient in Colorado. Third, we assessed the temporal dynamics of trait distributions and ecosystem net primary productivity using the Park Grass Experiment (PGE) from Rothamsted, UK. Lastly, to assess potential linkages with larger scale biogeographic gradients we review recent studies that assess shifts in trait distributions across large-scale biogeographic environmental gradients.

We primarily focus on assemblage variation in one trait, SLA, because it appears to vary more than other traits in eqn 11. Thus, we begin to assess predictions from TDT (Table 3) by substituting the mean SLA value for $\bar{b}_0$. If other traits in eqn 11 also vary or covary with each other across gradients, TDT would allow us to explore this as well.

For example, a shift in the mean community carbon use efficiency (or $c$) will lead to a decrease in NPP as observed (DeLucia *et al.*, 2007). Utilizing eqn. 11 we can now codify several additional TDT predictions based on SLA (see Table 3).

*Shifts in trait distributions and ecosystem measures across local abiotic gradients* – We tested several predictions generated by TDT (Table 1) with data collected along an elevational gradient in Colorado. We (Henderson, Sloat and Enquist) have measured community composition, ecosystem fluxes and traits of *all* individuals in several communities across an elevational gradient within subalpine communities near the Rocky Mountain Biological Laboratory (RMBL, Gunnison Co., CO, USA). Sites ranged from 2,460 m to 3,380 m and had similar slope, aspect and vegetation. The lowest elevation site is characterized as a semi-arid sagebrush scrub whereas subalpine meadow communities dominate at the higher elevations. Leaf traits were measured from every individual within a 1.2 x 1.2m plot. Measures of total ecosystem carbon production, community weighted SLA and variances were obtained by harvesting biomass and measuring total ecosystem carbon fluxes or Net Ecosystem Production, NEP (umol $CO_2$ $m^{-2}$ $sec^{-1}$). A more detailed listing of the methods used in our analyses is given in the Supplemental Document.

*Local Tests of TDT Predictions* - We next tested several predictions generated by TDT (Table 1) with data from the Park Grass Experiment (PGE) data from Rothamsted, UK. The PGE follows grassland plant community composition over a 140-yr period. Started in 1956, it is the oldest ecological experiment in the world (Silvertown *et al.*, 2006).

The dataset is unique as it allows us to assess community responses to an environmental driver - the experimental altering of soil nutrient availability. We use this dataset to assess how a change in the environment, *E*, in this case soil nutrients, differentially influences community composition and ecosystem function via the trait distribution, *C*(*z*). Within this experimental setup, the main environmental driver is a nutrient addition in the fertilized plot. We first focused on quantifying community SLA frequency distributions. However, as a more direct test of TDT, we also assessed two other key traits, plant height and seed size (see Supplemental Document). To approximate the trait distribution, *C*(*z*), we assigned species mean traits to species found within the PGE from the LEDA database (Kleyer *et al.*, 2008). A detailed listing of the methods used in our analyses is given in the Supplemental Document including background of the Park Grass Experiment.

## VII. Results

*Community trait shifts across local gradients* - Numerous studies have documented shifts in the traits of communities and assemblages across environmental gradients (Fonseca *et al.*, 2000; Ackerly, 2003; Choler, 2005; Swenson & Enquist, 2007). However, many studies generally calculate a species mean trait as part of a species list thus ignoring intra-specific variation. In contrast, several recent studies have measured traits within communities to assess community-level trait shifts (Gaucherand & Lavorel, 2007; Lavorel *et al.*, 2008; Albert *et al.*, 2010; Hulshof *et al.*, 2013). For example, Cornwell and Ackerly (2009) show that across a gradient of water availability, the community mean and intraspecific mean SLA significantly shifted such that drier environments have lower mean SLA (Fig. 4A).

*Shifts of community trait distributions across environmental gradients* - Data from our elevational gradient at the Rocky Mountain Biological Lab, Gothic CO (see Bryant *et al.*, 2008; Sides *et al.*, 2014; Sloat *et al.*, 2014), provide one of the first studies to measure the functional traits of every single individual within a given community (Fig 4B). Few studies have fully documented the community trait distribution by measuring trait values from every individual within the community.  In this system increasing elevation is associated with a decrease in temperature and increase in precipitation, and these changes drive the observed increase in SLA with elevation (Sides *et al.*, 2014). With increasing elevation, leaves have less structural durability, and lower life spans due to a shorter growing season but have higher photosynthetic rates (Enquist, Henderson, Sloat unpublished data ). According to TDT the elevational trend in SLA is due to a shift in the optimum trait value based on temperature and water availability, with a corresponding shift in the range of successful trait values.

Assessing shifts in inter- and interaspecific trait variation allows us to assess two central assumptions of TDT. First, trait distributions show directional shifts across gradients. According to Eqn. 10, for a given *E*, those individuals with phenotypes that are closer to the mean community value should have, on average, the highest growth rates. Previous studies along this same gradient and study site have documented a rapid turnover of species with elevation (Bryant *et al.*, 2008). Figure 4B indicates that the strong species diversity gradient is also reflected by a shift in traits. Note the range of trait variation is approximately 2-3 orders of magnitude. This range of trait values observed within communities in our gradient is approximately ½ of the fraction of the variation observed in SLA across the globe across all plants (Reich *et al.*, 1997). So, across the span of about ~25km distance between their study sites, we observe a significant fraction of the trait variation that is observed within species across the globe. As more studies document shifts in SLA across strong environmental gradients (such as elevation, flooding, soil water availability, disturbance), it is becoming clear that the magnitude of change can be nearly as large the global variation in the trait (Elser *et al.*, 2010; Violle *et al.*, 2012). These results suggest that more local studies of community trait distributions are reasonable proxies or natural laboratories for scaling up trait-based ecology across large global climate gradients as well as to predict future climate change scenarios.

Second, analyses from Sides et al. (2014) and Cornwell and Ackerly (2009) also provide a key assessment to a core assumption of TDT. According to TDT, for a given environment, *E*, if there is a mean optimal phenotype that maximizes growth rate given an environmental tradeoff (Fig. 2) then a external filters and/or selection/plasticity will then promote convergence of traits around this local optimal phenotype (Norberg et al. 2001; see also Violle et al., 2012). Both of these studies show that patterns of *intra*-specific mean trait shifts across an environmental gradient are in the *same* direction as the interspecific community shift across the gradient (see Fig.4A). In other words, intraspecific trait variation in local populations shift in the *same* direction as the community trait distribution.

This is consistent with the expectation that either selection and/ or phenotypic plasticity has resulted in individuals that adjust their phenotypes to better match an optimal phenotype within each community.

*Shifts of assemblage trait distributions across broad environmental gradients* - Across broad-scale geographic gradients, recent geographic trait mapping analyses from Swenson et al. (2012) and Šímová et al. (2014) show that the mean assemblage trait value of many plant functional traits vary directionally across biogeographic scales. Geographic variation in the mean tree assemblage SLA as well as tree size (height, a proxy for plant mass, $m$) shows significant shifts in both traits across gradients at the biogeographic scale (Fig 5). As is assumed in TDT, across a given environmental gradient, $E$, the mean community value, $C(z)$ or $C(b_0)$ will shift. Indeed, across North American, Šímová et al. find that the mean assemblage SLA is positively correlated with annual precipitation ($r^2$=0.539) but negatively related to annual temperature seasonality ($r^2$=-0.440) (see Table 2 in Šímová et al. (2014)).

*Building better models for variation in ecosystem function via the shape of trait distributions and Metabolic Scaling Theory* –

At the local scale measures of trait distributions associated with our theory in principle can be used to scale up to ecosystem function as well as to predict potential future community responses to climate change. In support of this we do find that a shift in the community specific leaf area is associated with a corresponding shift in net ecosystem productivity or carbon flux or Net Ecosystem Production – the net carbon uptake of the ecosystem (umol $CO_2$ m$^{-2}$ sec$^{-1}$). Focusing on the key trait SLA and variation in community biomass, $C_{Tot}$, to log transformed data, we fit a simplified version of the TDT scaling model $NEP \propto \langle CWM_{SLA} \rangle \cdot \langle CWV_{SLA} \rangle \cdot C_{Tot}^{b}$ using site as a factor and allowing the scaling exponent, $b$, to float. Here the values $\langle CWM_{SLA} \rangle$ and $\langle CWV_{SLA} \rangle$ are the community abundance weighted mean and variance in SLA respectively. The fitted model predicts 0.778 of the variation in net ecosystem production or NEP (df=22, F = 11.04, p < 0.0001, AIC = -24.39). Further, increasing variance, $\langle CWV_{SLA} \rangle$ decreases NEP. So, after for controlling for site level differences, communities with more variance in SLA had lower NEP. In contrast, variation in species diversity explains none of the variation NEP in this system ($r^2$=0.029, $p$ = 0.3617, AIC = 7.92). Additional analyses underscore the importance of both $\langle CWM_{SLA} \rangle$ and $\langle CWV_{SLA} \rangle$ on influencing variation NEP across this elevational gradient (see Supplemental Information). Together, these results support several key predictions of TDT - the shift in the mean of $C(b_0)$ is closely tied to environmental drivers and that shifts in the mean and variance of $C(b_0)$ is a primary driver of variation in community carbon flux (Fig 4C-D).

Across broad climatic gradients, recently Michaletz et al. (2014) utilized MST to predict variation in annual net primary productivity (NPP, grams of biomass per area per year). In support of MST, rates of growing season NPP scaled with total autotrophic biomass indistinguishable from the allometrically ideal value of 3/5 predicted value in eqn. 12. Their analysis support another prediction of TDT that controlling for scaled effects of total biomass, $C_{Tot}$, on NPP shows that shifts in $\bar{b}_0$, primarily because of shifts in <SLA>, will also shift variation in NPP (Fig. 7).

At larger biogeographic scales, assemblage trait maps such as Fig 5 could then be used to predict ecosystem functioning. At these larger geographic scales, if the distribution of SLA still reasonably approximate rates of biomass production, then, according to TDT, regions with high mean SLA and low variance should have the highest rates of instantaneous net primary production. In general, recent compilation of geographic variation in instantaneous rates of terrestrial ecosystem NPP from remotely sensed data indicate that areas with the highest instantaneous rates of NPP do generally correspond[2] to assemblages with high mean and low variance in SLA. However, according to eqn. 16 one should also control for total system biomass (which correlates with variation in tree height) as well as variation in the other traits that also can influence NPP. Nonetheless, the correspondence between biogeographic variation traits and predictions from TDT is a promising future direction.

*Temporal trait shifts across fertilization gradients* – The results from the long-term dynamics and fertilization experiment from Rothamsted are given in Tables 2-3 as well as Figure 6 and S3-S5. Within the Rothamsted dataset, all of the traits studied, the biomass-weighted distribution of the trait specific leaf area, SLA, was associated with the most prominent shifts in the central moments of communities in response to fertilization (Fig S1, Table 2; see also Supplemental Document). For both fertilized and control plots, all four moments of the community SLA distribution changed significantly. The overall effect of fertilization on the community trait distribution is consistent with fertilization differentially favoring certain phenotypes (Chapin & Shaver, 1985; Suding *et al.*, 2005) and a replacement of slower growing species with faster growing species (Grime & Hunt, 1975; Chapin, 1980; Knops & Reinhart, 2000). Similar to past findings, across all plots, the community mean SLA increased (see Knops & Reinhart, 2000; Craine *et al.*, 2001) and the variance decreased. The directional community trait shift is reflected in increased skewness values (differing from zero). Fertilization also led to a shift in the kurtosis but only for SLA values. Specifically, SLA shifted from negative kurtosis values to zero or positive kurtosis values suggestive that fertilization increased rates of competitive exclusion of suboptimal trait values leading to a more peaked trait distribution. Intriguingly, the direction and rate of change for fertilized versus control plots differed in sign and magnitude indicating that the trait distributions of control and experimental plots steadily diverged over time.

---

[2] http://daac.ornl.gov/NPP/npp_home.shtml#

The Park Grass Experiment supports several predictions from TDT. First, TDT predicts that a shift in an environmental driver (fertilization in this case) should primarily be seen as a shift in traits associated with growth rate. Of all of the traits assessed, SLA showed the strongest shifts over time (Table 1). The other traits showed relatively little to no change over time. The disproportionate shift in SLA is consistent to expectations from TDT as SLA is the only trait directly linked to the growth function, $f(b_0)$. Second, consistent with a shifting in an optimal phenotype, the skewness of the fertilized plot increased over time. Third, consistent with a productivity-trait variance tradeoff, the annual net primary productivity (NPP) was positively correlated with community mean and kurtosis of the SLA distribution but negatively correlated with variance in SLA (Fig. 6; Table 3), Lastly, shifts in the trait distribution are more closely tied to NPP than species richness (see Table 3). We observe a weak to negative correlation between species richness and NPP (Fig. 6A; Table S1), which is opposite to a hypothesized positive relationship predicted solely from Biodiversity and Ecosystem Functioning (BEF) theory on species richness (Tilman *et al.*, 1997).

**Discussion**

We have shown that Trait Driver Theory or TDT can formalize numerous assumptions and approaches in trait-based ecology. We provide examples of how this can be done for several different biodiversity hypotheses in terms of the dispersion of traits (Table 1 & 3). We further argue that ecological theories need to move beyond species richness and be recast in terms of organismal performance via functional traits. As a result, TDT offers an alternative framework to the standard taxonomic approach for linking biodiversity and ecosystem functioning, where primacy has been placed on the importance of species richness. TDT instead focuses on the importance of 'trait diversity' via the shape of the trait distribution of individuals and shared performance currencies (e.g., growth). Because TDT incorporates intraspecific variation, it necessarily includes natural selection as a process that shapes the trait distribution. By incorporating interspecific trait variation, it also includes "selective" processes at higher levels of organization within the community, such as species sorting (see also Shipley et al. (2010)). We show that using TDT to analyze these processes leads to several predictions opposite to predictions made by Biodiversity Ecosystem Function theory (Tilman *et al.*, 1997; Naeem & Wright, 2003; see also Table 3), it also offers a useful alternative hypothesis by which to assess the linkage between 'diversity' (whether measured by species richness or via the trait distribution) and ecosystem functioning.

TDT also offers a predictive framework for management. Increasingly, trait-based approaches to management have shown that a focus on trait shifts due to land use, as well as management and agricultural practice, can yield deeper insight into the processes of concern to managers (Garnier & Navas, 2012). For example, biomass production, the timings of peak production and plant digestibility, response time to disturbance can be predicted from the shape of the community trait distribution as well as many of the plant traits underlying our general growth equation (see studies and references listed in Garnier & Navas, 2012).

Our analyses find that none of the central-moments of the trait distribution in the Park Grass dataset are correlated with species richness (see Table S1 in the Supplemental Document). Indeed, species richness does not appear to be a reliable proxy for how the diversity of phenotypes and trait distributions respond to environmental change. Further, across large biogeographic gradients, recent studies have found that total functional trait space and dispersion are unrelated to species richness (Safi *et al.*, 2011; Lamanna *et al.*, 2014; Šímová *et al.*, 2014). The potential for improved predictions using trait distributions linked to metabolic scaling fundamentally comes down to the increase in information contained in traits that is not necessarily present in a species richness-based approach (Tilman *et al.*, 1997) or even a phylogenetic approach to community ecology (Cavender-Bares *et al.*, 2009). Although TDT is based on simplifying assumptions, it helps to better connect and scale trait-based ecology and MST with large-scale ecology and biogeography. It integrates and builds upon prior work that (i) developed highly mechanistic trait-based models (Norberg *et al.*, 2001; Zhang, 2013) and other work that predicted how individual growth rates change across scale with size and temperature (Enquist *et al.*, 2007a). TDT thus enables ecological theories to be 'scaled up' to predict and test the consequences of how organismal response to climate change will ramify at the community and ecosystem levels across both local- and large-scale gradients in geography (space) and fluctuations in time (climate change).

It is becoming clear that multiple assembly processes—abiotic filtering, biological enemies, competition, facilitation, below ground competition—likely operate simultaneously and at differing scales to structure communities and larger scale species assemblages (Grime, 2006; Cavender-Bares *et al.*, 2009; Swenson & Enquist, 2009; Mayfield & Levine, 2010). The result is that the distribution of some traits may be more over- or under-dispersed than others. Future work should better link empirical data with theory to test the community and ecosystem responses when multiple trait drivers influence trait variance and when trait optima are strongly influenced by differing levels of ecological interactions (competition, predation, mutualism etc.). Our work has primarily focused on the traits and environmental drivers that underlie growth rate. Indeed, refining and extending TDT will also require better identification of the above and below ground traits that influence growth rate. Lastly, analysis of the shift in the mean and variance of assemblage values of many plant functional traits as well as stand biomass across biogeographic scales may provide the necessary basis to predict ecosystem functioning across large scales. Focusing on variation in specific leaf area, SLA, TDT predicts that the distribution of SLA will influence rates of biomass production. Correcting for the effects of stand biomass (see Eqn. 8), assemblages with high mean SLA and low variance should have the highest rates of instantaneous net primary production. In general, recent compilation of geographic variation in instantaneous rates of NPP from remotely sensed data indicate that areas with the highest instantaneous rates of NPP generally do correspond (Zhao *et al.*, 2005; Zhao & Running, 2010) to assemblages with high mean and low variance in SLA regions identified by Swenson et al. (2012) and Simova et al. (2014) . TDT also predicts that while regions with relatively lower trait variance will be more sensitive to rapid directional climate change, assemblages with greater variance, however, would be expected to more closely track climate change. Future tests of TDT at the scale of global ecology should more formally assess the predictions of TDT at this scale by assessing the specific relationships between the trait distribution, vegetation biomass, and possible covariation of other traits.

The TDT prediction of an inverse relationship between trait variance and production is not necessarily in conflict with either "positive species complementarity"—niche partitioning allow species to capture more resources in ways that are complementary in both space and time (Tilman, 1999) or "transgressive overyielding" where species use resources in ways that are complementary in space or time to stably coexist with one another so that more diverse communities capture a greater fraction of available resources and produce more biomass than even their most productive species (see Tilman et al. 1997). TDT needs to be reconciled with these ideas because recent studies confirmed that within biodiversity experiments, "positive species complementarity" does enhance ecosystem productivity (Cardinale *et al.*, 2007). Given TDT, a natural question is how are trait distributions modified when complementarity effects are strong? Effects such as complementarity can be incorporated into TDT growth functions as explained in Savage et al. (2007).

In sum, we have argued that more powerful tests of biodiversity theories need to move beyond species richness and explicitly focus on mechanisms generating diversity via trait composition and diversity via the shape of trait distributions. The rise of trait-based ecology has led to an increased focus on the distribution and dynamics of traits across broad geographic and climatic gradients and how these distributions influence ecosystem function. However, a trait-based ecology that is explicitly formulated to apply across different scales (e.g., species that differ in size) and gradients (e.g., environmental temperatures) has yet to be articulated. The Trait Drivers Theory, TDT, presented here is a formalization of essential steps for mechanistically linking and scaling functional traits for individuals with the dynamics of ecological communities and ecosystem functioning. This TDT approach builds upon and complements existing trait-based approaches in ecology (e.g. Grime, 1998; Lavorel & Garnier, 2002; Kraft *et al.*, 2008; Suding *et al.*, 2008a). It is appealing because it can connect individual physiology and traits with ecosystem dynamics and how both respond to climate change (Suding *et al.*, 2008b), geographic gradients, and differing ecological processes (e.g. niche versus neutral see Weiher et al. 2011). Given the increasing ability to remotely sense numerous traits of terrestrial vegetation (Doughty *et al.*, 2011; Asner *et al.*, 2014) and the increasing access to both plant and animal trait data (Kattge *et al.*, 2011), TDT and its elaborations are ripe for providing empirically-grounded, mechanistic models of ecosystem dynamics from local to large scales.

**Acknowledgments**

We thank Mark Westoby who provided enthusiasm and input during the writing of initial drafts of this paper. We also thank the ARC-NZ Vegetation Network, working group 36, led by VMS, for their generosity and support in bringing us all together to meet and to initiate this collaboration and paper. We thank other members of working group 36, especially Christine Lamanna, Tony Dell, Mick McCarthy, and Graham D. Farquhar, who helped us solidify our central arguments. C.V. was supported by a Marie Curie International Outgoing Fellowship within the 7th European Community Framework Program (DiversiTraits project, no. 221060). B.J.E. and V.M.S. were supported by an NSF ATB award and BJE by an NSF Macrosystems award. C.T.W. was supported by NSF Grant DEB-0618097. JN was supported by the Swedish research council and FORMAS grant EKOKLIM. Lastly, we thank the Park Grass

Table 1. Summary of the core predictions from Trait Drivers Theory for how the different central moments of the trait distribution will respond to differing biotic and abiotic forces and how they will then in turn influence community dynamics and ecosystem functioning. Trait Driver Theory (TDT) can incorporate each of these forces via the shape of the trait biomass distribution, $C(z)$, to then make specific predictions for how each can drive the dynamics of $C(z)$ and ecosystem functioning (see text). Parameterizing predictions for specific cases depends upon the traits that affect the growth rate, $f$.

| Moment of Community Trait Distribution, $C(z)$ | Predictions for rate of community response to a changing environment | Predicted Ecosystem Effects |
|---|---|---|
| I. Mean | (a) Will shift if environmental change alters value of $z_{opt}$ and time scales are not too rapid and oscillatory<br>(b) Lags $z_{opt}$ by an amount that depends on rate of change in environment, rates of immigration, and the forces that influence the variance. | (i) Will shift productivity according to form of growth equation, $f$. |
| II. Variance | (a) Decreases with strong abiotic filtering.<br>(b) Decreases with strong rates of competitive exclusion<br>(c) Can increase with immigration and reduced competition.<br>(d) Under neutral theory, if no immigration or mutation, variance will decrease over time so as to decrease response abilities over time. | (i) Increased variance implies lower productivity for fixed or stable environment.<br>(ii) Increased variance accelerates community response to environmental changes.<br>(iii) Increased variance will lead to increased stability of ecosystem functioning by reducing the lag of $\bar{z}$ and $z_{opt}$ in varying environments. |
| III. Skewness | (a) Skewness values > or < 0 can reflect a lag between $\bar{z}$ and $z_{opt}$ and a rapidly changing community due to an environmental driver or extreme limit to a trait value.<br>(b) Increases in skewness can indicate a response to rapid environmental changes or the importance of rare species advantages in local coexistence. | (i) Depending upon kurtosis and variance value, productivity should be reduced compared with a community with similar variance but skewness equal to zero. |
| IV. Kurtosis | (a) Positive kurtosis reflects competitive exclusion or other types of biotic exclusion<br>(b) Kurtosis close to -1.2 reflects a uniform distribution consistent with uniform niche partitioning.<br>(c) More negative values could reflect the coexistence of contrasting ecological strategies, recent or sudden environmental change. | (i) If the trait mean equals $z_{opt}$, forces that decrease kurtosis will decrease producitity while forces that increase kurtosis will increase productivity.<br>(ii) In a varying environment, greater kurtosis will lead to increased stability of ecosystem functioning by reducing the lag of $\bar{z}$ and $z_{opt}$. |

**Table 2.** Observed temporal changes in the central moments of the community trait distribution $C(z)$ in the Park Grass Experiment. Predictions from Trait Driver Theory correspond to the cells indicated in Table 1. As an estimate of the central-moments of $C(z)$ we estimated the community weighted values for the mean, variance, skewness and kurtosis (CWM, CWV, CWS, and CWK, respectively). Values are the Pearson product-moment correlations, $r$, are for the three main traits investigated, specific leaf area or SLA ($a_L/m_L$), adult height, and seed size. In accordance with TDT, for the trait SLA, the distributions of the fertilized and unfertilized plots have diverged for all trait moments. Further, in accordance with TDT, increasing fertilization leads to an increasing skew of the SLA distribution, but not for seed size and reproductive height, traits that do not directly underlie the growth equation, indicating that fertilization is a strong environmental driver that influences plant growth. Correlations are for both unfertilized: plot 2 and fertilized: plot 16. *, $P < 0.05$ ; **, $P < 0.01$ ; ***, $P < 0.001$ ; ns., not significant; (-), negative relationship; (+) positive relationship. Observed shifts in the central moments of SLA are generally in accordance to predictions from TDT (see Table 1 and text for details).

| | Moment of Community Trait Biomass Distribution, $C(z)$ | Control (Unfertilized) | Fertilized | Corresponding TDT Predictions in Table 1 |
|---|---|---|---|---|
| SLA | Mean (CWM) | 0.65*** (-) | 0.40** (+) | See I.a and I.b |
| | Variance (CWV) | 0.48** (+) | 0.26* (-) | See II. a-e |
| | Skewness (CWS) | 0.13$^{ns.}$ | 0.29* (-) | See III a-b |
| | Kurtosis (CWK) | 0.69*** (-) | 0.10$^{ns.}$ | See IV a-b |
| Height | Mean (CWM) | 0.61** (-) | 0.04$^{ns.}$ | See I.a and I.b |
| | Variance (CWV) | 0.27* (-) | 0.51** (+) | See II. a-e |
| | Skewness (CWS) | 0.17$^{ns.}$ | 0.02$^{ns.}$ | See III a-b |
| | Kurtosis (CWK) | 0.38** (+) | 0.20$^{ns.}$ | See IV a-b |
| Seed mass | Mean (CWM) | 0.21$^{ns.}$ | 0.01$^{ns.}$ | See I.a and I.b |
| | Variance (CWV) | 0.38** (-) | 0.01$^{ns.}$ | See II. a-e |
| | Skewness (CWS) | 0.56** (-) | 0.00$^{ns.}$ | See III a-b |
| | Kurtosis (CWK) | 0.31* (-) | 0.01$^{ns.}$ | See IV a-b |

**Table 3.** Correlations between trait moments and variation in annual Net Primary Productivity for the Park Grass experiment. Also listed are the predicted signs of the correlations made from Trait Driver Theory (see text and Table 1) and from Biodiversity-Ecosystem Functioning Theory (assuming the theory of Tilman et al. 1997 where variability in trait SLA is a good proxy for variation in species richness via niche or trait space). Note, both theories make opposite predictions for the signs of the correlations. NP = no specific prediction is made as such predictions would depend upon specifics of system. As an estimate of the central-moments of $C(z)$ we estimated the community biomass weighted values for the mean, variance, skewness and kurtosis (*CWM, CWV, CWS*, and *CWK*, respectively).

|     | Central Moment of Community Trait Distribution | Observed correlations | Predicted Response – Trait Driver Theory | Predicted Response – Biodiversity Theory |
|-----|---|---|---|---|
| SLA | Mean (CWM) | 0.71*** (+) | + | NP |
| SLA | Variance (CWV) | 0.45*** (-) | - | + |
| SLA | Skewness (CWS) | 0.28*** (+) | NP | NP |
| SLA | Kurtosis (CWK) | 0.19*** (+) | + | - |

\*, $P < 0.05$ ; \*\*, $P < 0.01$ ; \*\*\*, $P < 0.001$ ; ns., not significant
(-), negative relationship; (+) positive relationship

**Figure Legends**

**Figure 1.** Path diagram representing the relationships of phenotypic traits, *z*, performance measures, *f*, and fitness, *w*. The arrows on the left are correlations between traits. The coefficients *β* represent the correlation coefficients between traits, *z*, functions, *f*, and ultimately fitness, *w*. Note, performance measures include growth rate as well as survivorship and reproductive rates. Metabolic scaling theory explicitly links traits to these performance functions. Figure modified from Kingsolver and Huey (2003) modified from Arnold (1983)

**Figure 2.** Graphical representation of two central assumptions of Trait Drivers Theory. Each curve represents positive performance (here shown as the growth function *f*) or growth rates of individuals characterized by a unique trait value, *z*. Each color then indicates how a given trait value then translates to variation in organismal growth rate across an environmental gradient, *E*. Because of a trade off between traits each trait has an optimal environment where growth is maximized, each trait exhibits a unimodal response to an environmental gradient, *E*. At different points along an environmental gradient different traits are characterized by the highest growth rate, *f*. Also, species can differ in the width of their performance curves, $\sigma_z^2$. Here in two communities, A and B, although several trait values can achieve positive growth in both A and B (for example in community A, the black, red, yellow, and grey trait all can grow in that location but within A there is one of t trait, the black trait, that has the highest growth rate). This trait value then is the optimal trait value, given the potential species pool, for that community. Note, in this example growth rates are only influenced by *E*. However, extensions of TDT (see text) can assess how the growth rate of a given trait value then is influenced by the presence and dominance of other trait values. Such trait-trait interactions could then modify the shape and breadth of each growth curve.

**Figure 3.** Conceptual diagram linking changes in optimal traits in changing environments with the frequency distributions for that trait in historic versus current environments. In this example, the optimal trait value (dotted line) has shifted to the right. Individual performance (growth, fitness, etc.) is highest at the optimal trait expression in an environment. Note, because the response of an assemblage cannot be instantaneous, the trait distribution has an increased skew. Further, since the highest trait frequencies are not yet at the optimal trait expression, the average performance in the current environment is lower than in the historic environment.

**Figure 4.** Two examples of shifts in community trait distributions across gradients. **A.** Data from Cornwell and Ackerly (2009) showing shifts in the community mean and intraspecific mean value of the trait distribution for leaf specific area or SLA. Dashed lines represent least-squares fits for a given species and show the change in the population mean *intra* specific variation across the gradient in soil water. Solid points and the solid line show the least-squares regression for the arithmetic mean community or interspecific value for SLA. Note, consistent with the assumption of a shift in an 'optimal phenotype' across environmental gradients, intraspecific trait shifts are in the same direction as the community or interspecific shift **B.** Data from our Colorado elevational gradient showing the probability density distributions of SLA based on all individuals in a 1.3 x 1.3m plot at five sites along an elevational gradient. The number of individuals at each site is: 2468m = 234, 2710m = 639, 2815m = 938, 3155m = 282, 3380m = 160. Across the elevational gradient the community trait distribution of all individuals significantly shifts with changes in the mean community trait and variance. In C and D, for the Colorado plots we assess how changes in the community weighted mean and variance of five plots within each site contributes to variation in net ecosystem production of $CO_2$ (NEP). These plots are partial residual plots showing linearization of NEP relationships with the community weighted values of mean SLA and variance in SLA. As predicted by TDT ecosystem carbon flux is positively related to a shift in mean SLA but negatively related to an increase in community trait variance. In contrast, variation in species diversity explains none of the variation NEP in this system.

**Figure 5.** Correlations between annual net primary productivity (NPP) and (**A**) species diversity per plot; (**B**) the plot community biomass-weighted mean, *CWM*, of the SLA distribution; (**C**) the plot community biomass-weighted variance, *CWV*, of the SLA distribution for the four selected plots from the Park Grass experiment. Each dot represents the annual aboveground biomass production of a given plot in a given year. Compare these relationships to the predictions with Trait Driver Theory and Biodiversity and Ecosystem Functioning Theory (Table 3). Note that there is a negative correlation between NPP and diversity while the opposite is predicted by classical biodiversity-ecosystem functioning theory. However, in accordance with TDT, there is a *negative* relationship between the variance of the abundance weighted community trait distribution and a *positive* relationship between mean weighted trait and NPP. These results show a direct linkage between a critical trait that influences plant growth, forces that influence the shape of the trait distribution, and how both changes in trait mean and variance then shape ecosystem functioning.

**Figure 6.** In support of Trait Drivers Theory – woody plant net primary production scales with total assemblage biomass. Further, for a given amount of biomass, NPP will be modified by shifts in the mean community NPP. However, TDT NPP will also be modified by the assemblage variance in SLA. Figure modified from Michaletz *et al.* 2014.

**Figure 7.** Biogeographic variation in the **A.** mean species assemblage functional trait leaf specific area (SLA, $m^2 g^{-1}$) and plant height (m) and **B.** the standardized effect size (SES) of the mean trait values across woody plant communities across North America. The mean SLA is lowest in the low elevation-latitude tropical forests and tends to be highest in temperate forest and grassland regions. Across broad geographic gradients both the mean and the variance significantly vary. According to Trait Driver Theory – areas that correspond to high mean SLA and larger standing stocks of biomass (corresponding to taller mean heights) should have the highest rates of instantaneous net primary production, NPP. However, NPP will also be modified by the assemblage variance in SLA. Figure modified from Simova et al. (2014).

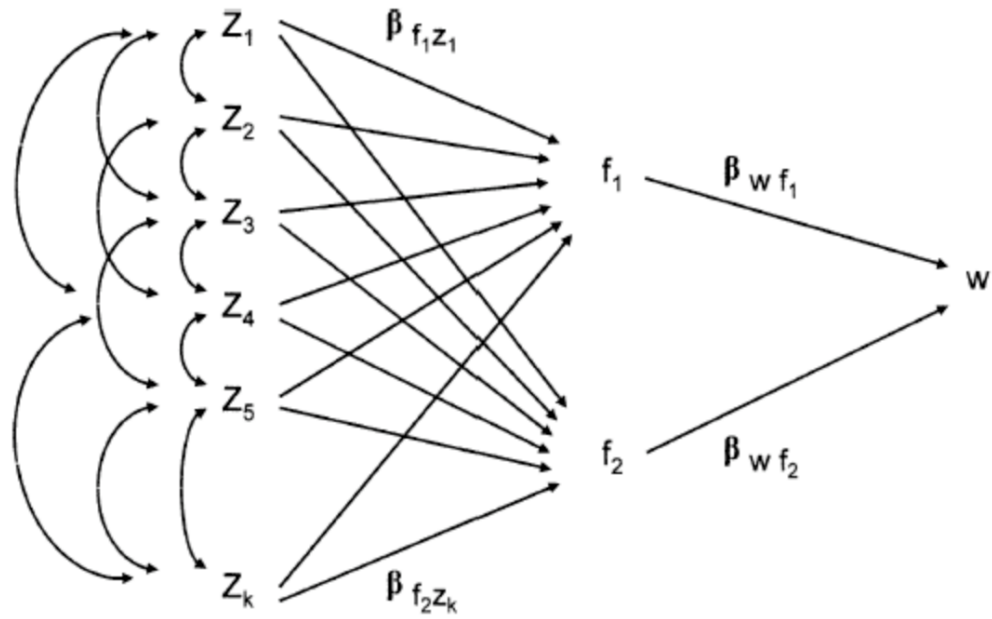

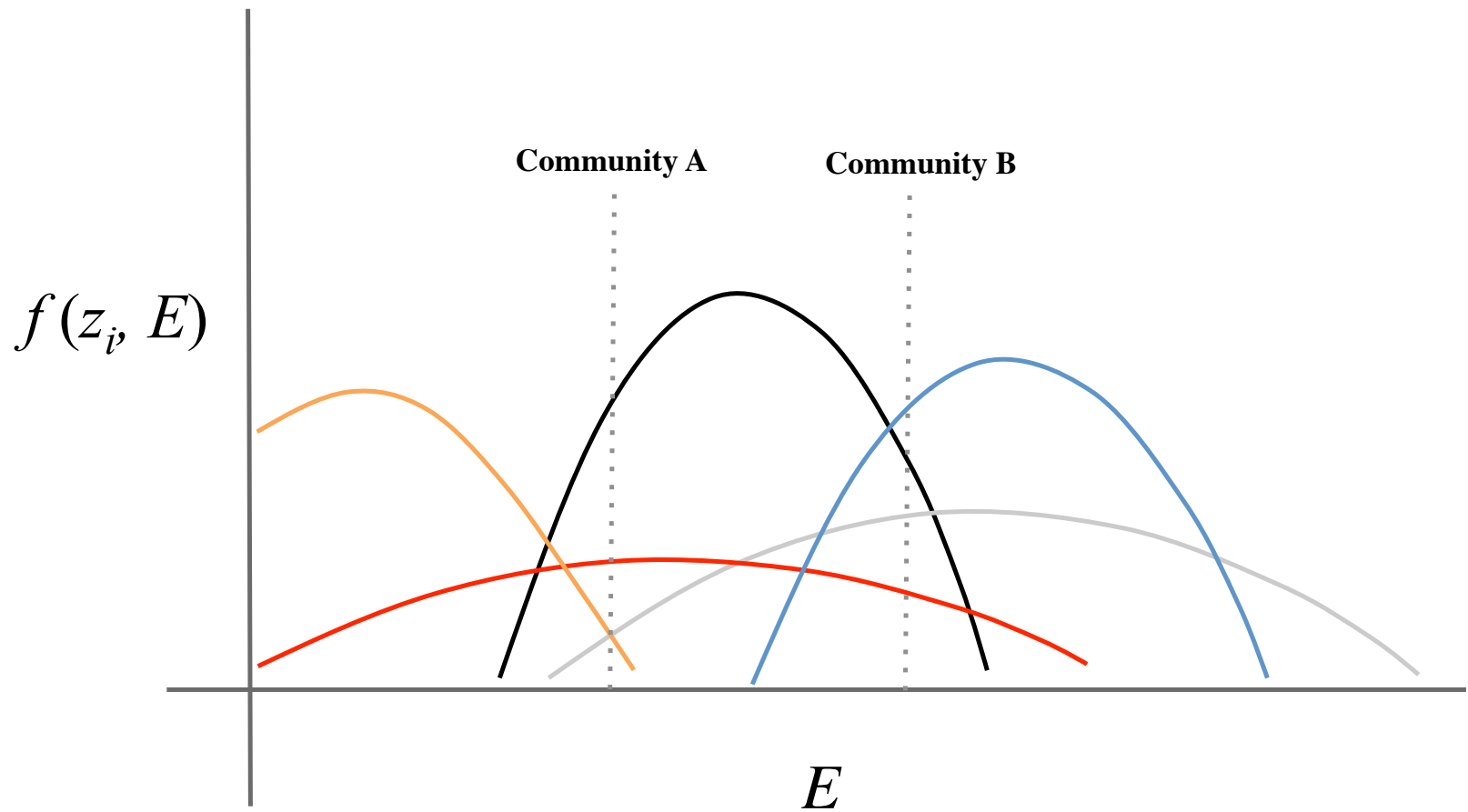

**Environment**

**Trait distribution:** Frequency of a trait expressed for all individuals (across all species) within a community

**Individual performance:** Traits related to growth rate, acquiring and allocating resources, etc. and ultimately fitness.

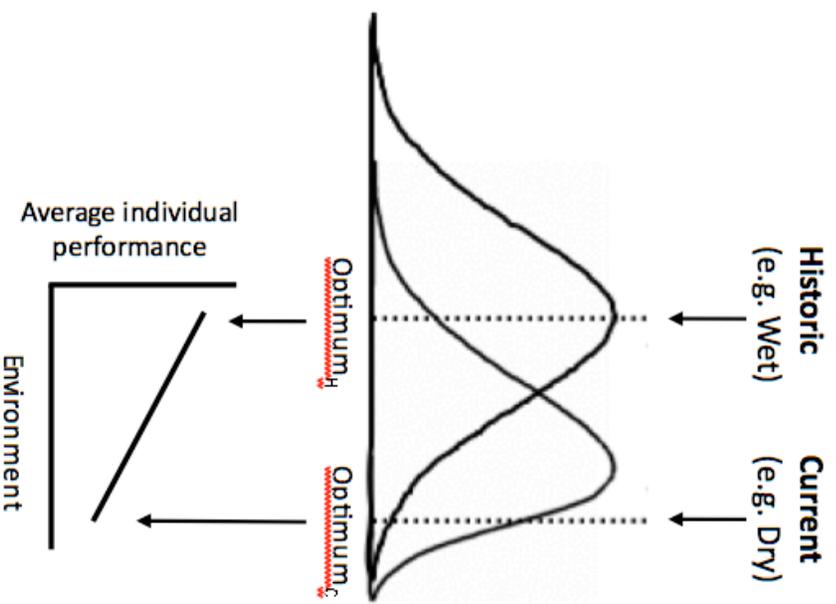

Historic (e.g. Wet)

Current (e.g. Dry)

Optimum<sub>H</sub> 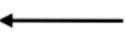

Optimum<sub>C</sub> 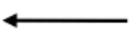

Average individual performance

Environment

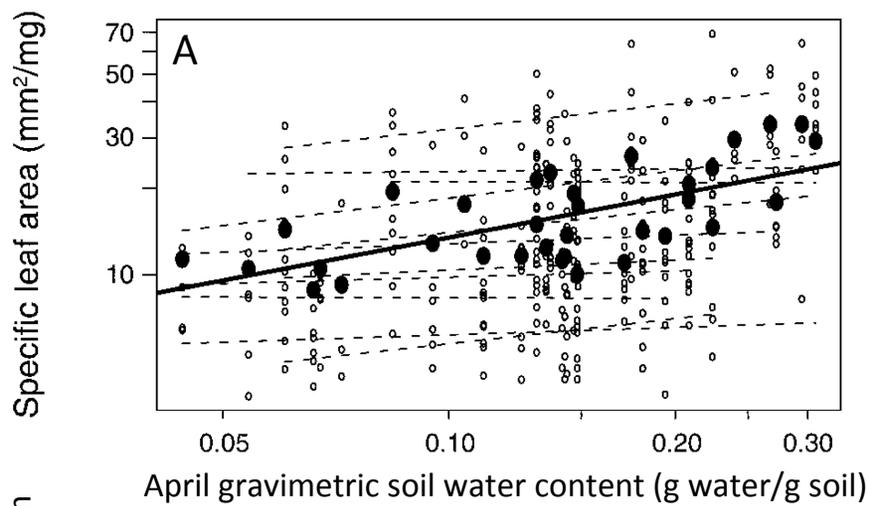
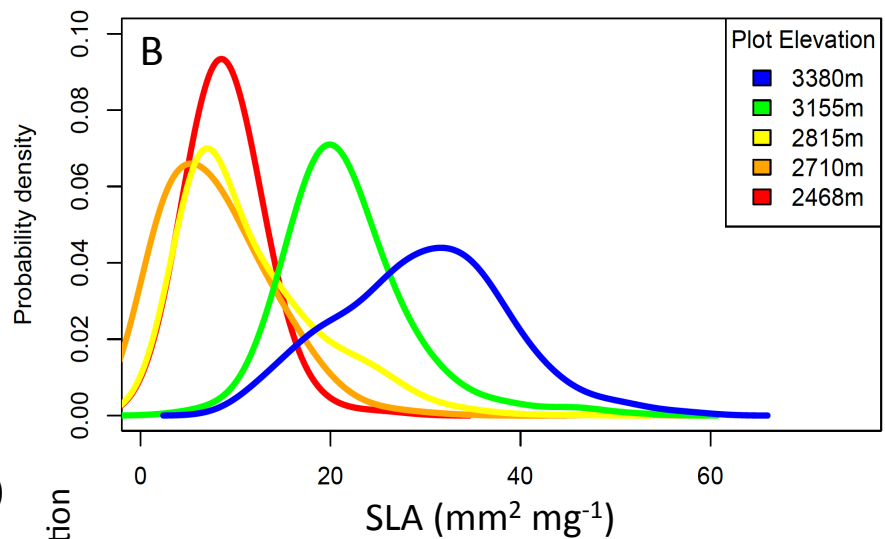
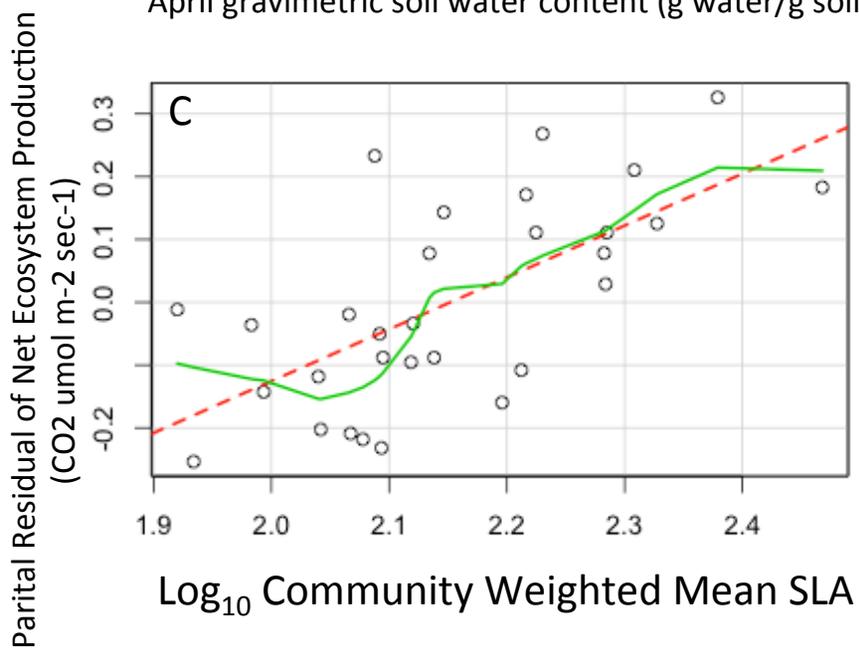
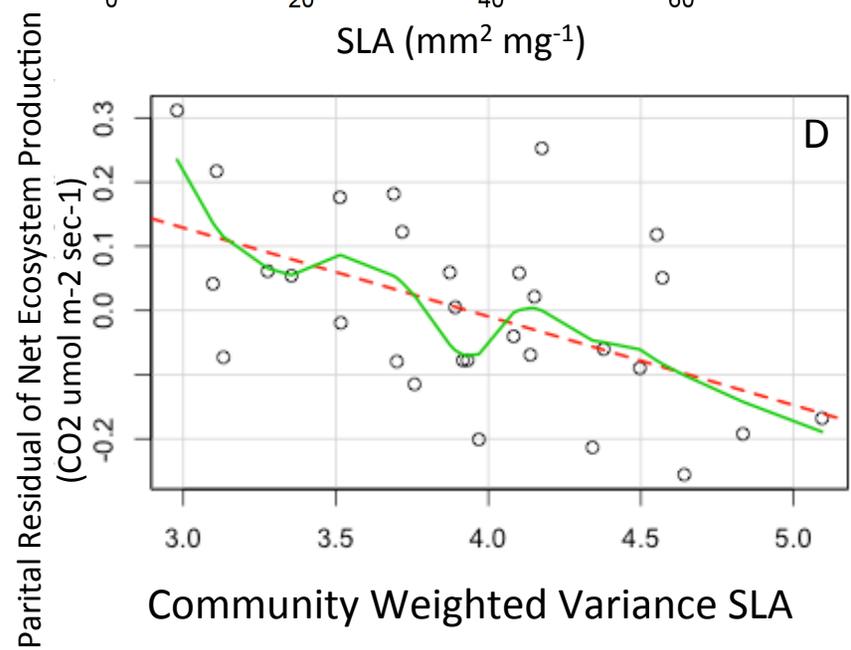

1
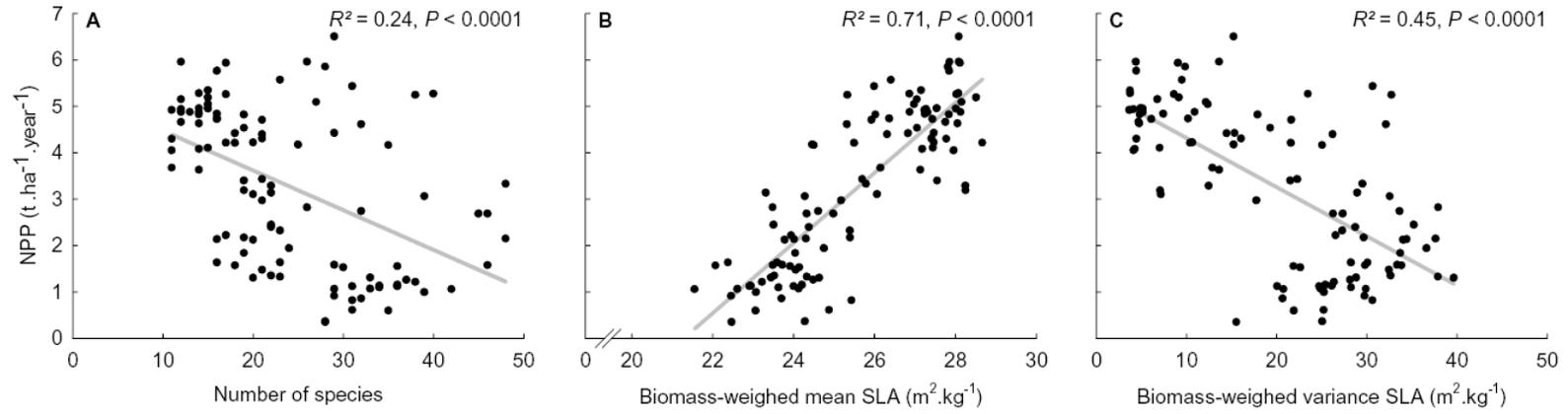
2

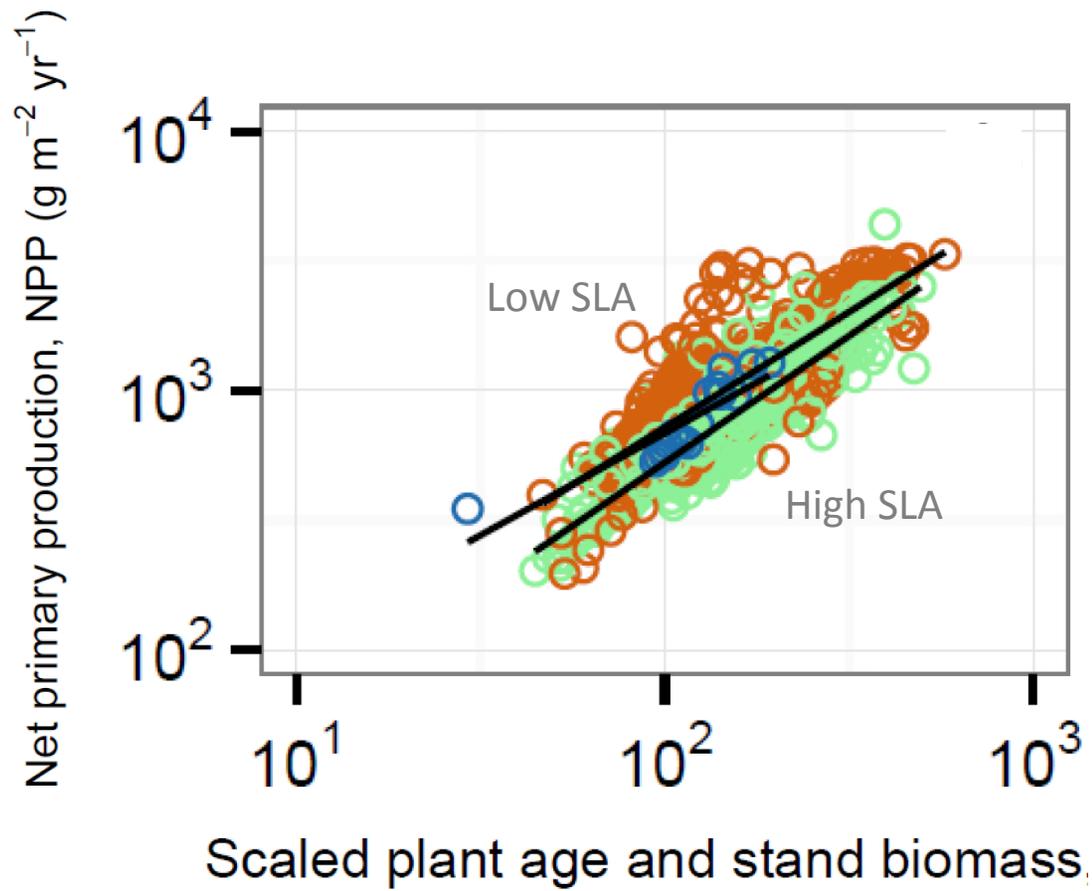

Mean Assemblage Trait Value

Trait Value Standardized Effect Size (SES)

Plant Size (height)

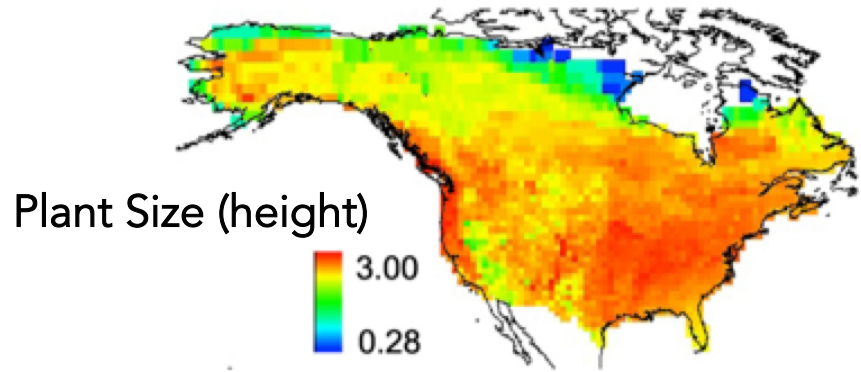
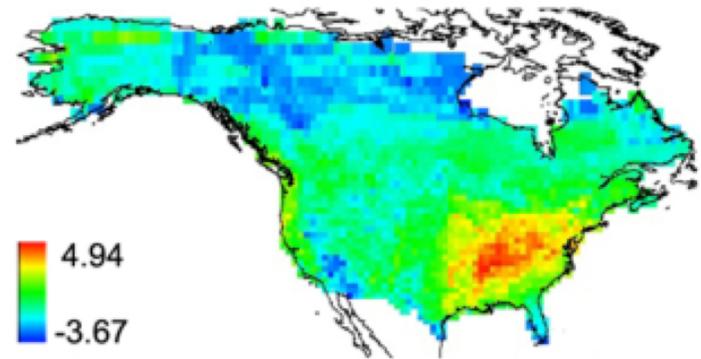

Specific Leaf Area (SLA)

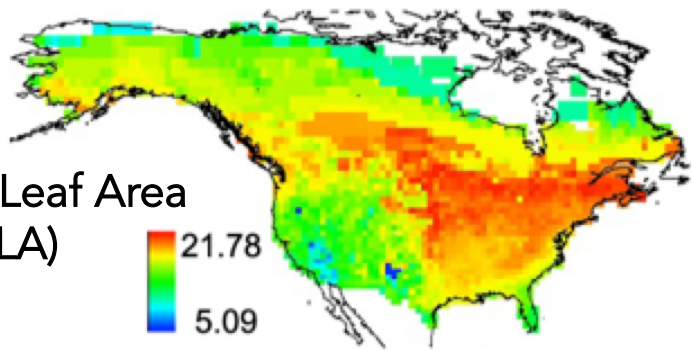
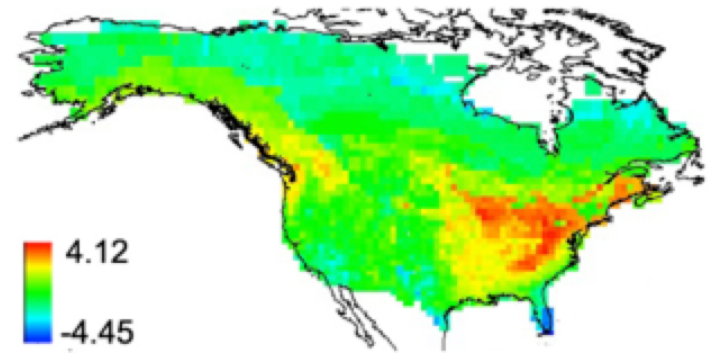

**Supplementary Document**

**I. Using TDT to recast ecological hypotheses**

**I. Using TDT to recast ecological hypotheses**

A unique attribute of TDT is that several differing ecological theories can now be recast in terms of how they influence the shape of trait frequency distributions ($C(z)$). Each of these theories make differing hypotheses that influence the shape of trait distributions and the functioning of ecosystems:

*(i) H1: Phenotype-environment matching–* This hypothesis states that species are more successful in different parts of the landscape because individuals have different trait values across space that are better adapted to local features of that space, such that the mean 'phenotype' matches variation in the local environment (Westoby & Wright, 2006). This distinction builds upon observations that stem back to Schimper (Schimper, 1898; 1903) that form the foundation for understanding changes in fitness and functional traits and species composition across gradients (Levins, 1968; Westoby & Wright, 2006). A prediction of this hypothesis is that, either due to convergent evolution or abiotic filtering of relevant traits, similar environments should be dominated by species with similar trait values (see discussion in Karr & James, 1975; Mooney, 1977; Orians & Paine, 1983).

*(ii) H2: The competitive-ability hierarchy hypothesis –* This hypothesis states that the strength of competition between individuals is driven by the distance between individuals as measured according to their functional traits (Freckleton & Watkinson (2001)). The competitive-ability hierarchy hypothesis leads to opposite predictions than the niche-based competition-trait similarity and competition-relatedness hypotheses, see below (Mayfield & Levine 2010). Here, the resulting competitive hierarchy will cause in a reduction in the trait variance over time and increased functional clustering because individuals that share a trait value will outcompete individuals with different trait values (Freckleton & Watkinson, 2001; Mayfield & Levine, 2010; Kunstler *et al.*, 2011).

*(iii) H2: Abiotic filtering -* The importance of local abiotic forces are reflected in the community trait range and variance – Abiotic filtering hypothesis states that increasingly more stressful environments will limit the range and variance influence. This hypothesis was formalized by Keddy and colleagues (Keddy, 1992; Weiher & Keddy, 1999; Kraft & Ackerly, 2010). Similarly, on ecological time scales, due to phenotype-environment matching, increasingly more stressful environments, E, will increasingly restrict the range of trait values that could co-occur within a given environment. This 'trait filtering hypothesis' states that the abiotic environment filters trait values so as to limit the variance and range of the trait distribution. This hypothesis, which can be seen as an ecological scale version of the 'favorability' hypothesis of Terborgh (Terborgh, 1973), predicts that more physiologically stressful environments (frost, high salinity, drought, etc.) should place especially rigid filters on the types of phenotypes (i.e., traits) that can survive and potentially co-occur (e.g. (Kraft *et al.*, 2008)). Note, as discussed above (ii) and below (iii), the variance and range of a trait distribution *C(z)* is also influenced by biotic forces. Similarly, repeated disturbance or environmental variability may minimize local interactions and could also increase community trait variance ((Grime, 2006).

*(iv) H3: Strength of local biotic forces is revealed via trait variance and kurtosis*-
Differing biotic community assembly hypotheses can differentially influence the spacing of trait values within the range of filtered phenotypes. As stated in (ii), competition for a common limiting resource would ultimately lead to competitive exclusion (e.g see Tilman, 1982) resulting in a convergence of 'superior competitor' phenotypes (Abrams & Chen, 2002; Savage *et al.*, 2007; Mayfield & Levine, 2010). This convergence would be reflected by decreasing variance and an increase in 'peakedness' of the trait distribution or an increase in positive kurtosis (Navas & Violle, 2009). In contract, according to Chesson (2000), if traits map onto niche differences, increased niche (trait) differentiation will lead to increasing coexistence of individuals with differing traits. These classical niche partitioning models predict that competition will limit functional (trait) similarity (MacArthur & Levins, 1967)and thus increase in the spacing between co-occuring phenotypes phenotypes (see MacArthur, 1958; Diamond, 1975). Similarly, biological enemies (Kraft & Ackerly, 2010), facilitation (Brooker *et al.*, 2008), and frequent disturbance (Grime, 1998) can maintain trait diversity (e.g. an over-dispersion of phenotypes). Niche packing models result in either a broader or evenly dispersed trait distribution (high variance) or even a multimodal trait (negative kurtosis) distribution.

*(v) H4: Assessing neutral forces via the shape of local and regional trait distributions* –
An alternative hypothesis to (ii) and (iii) is that local communities are primarily structured by stochastic dispersal, drift, and dispersal limitation (Hubbell, 2001). Such a neutral scenario would predict on average, for traits not associated with dispersal, little to no difference in the shape of the community trait distribution when sampled across differing spatial scales. Further, for traits not associated with dispersal ability, there should be no relationship between trait distribution and changes in the environment.

**II. Integrating Metabolic Scaling Theory into Trait Drivers Theory**

Within Trait Driver Theory (TDT), the total biomass associated with a trait, *z*, is denoted by *C(z)*. This notation avoids ever needing to account for individual mass, *M*, or number of individuals with mass. In contrast, the growth equations Metabolic Scaling Theory are phrased in terms of individual mass, *M*, where organismal growth rate, *dM/dt* is given by

$$\frac{dM}{dt} = b_0(z) M^{3/4}$$

(1)

where $b_0(z)$ is a coefficient that depends on a single or set (meaning *z* is a vector) set of traits.

To integrate MST and TDT we first recognize that, *C(z)*, the biomass associated with trait *z* can be expressed as $C(z) = \int dM \, C(z, M) = \int dM \, N(z, M) M$, where *C(z,M)* is the mass density of individuals with both trait value *z* and individual mass *M,* while *N(z,M)* is the number density of individuals that have both trait value *z* and individual mass *M*. In this expression, we have integrated over all possible values of mass, *M*, so that have the total biomass of *all* individuals with trait $z$. Furthermore, note that integrating this over all traits, $z$, gives the total biomass, $C_{TOT} = \int dz\, C(z) = \int dz \int dM\, N(z,M)M$. Consequently, we can multiply both sides of Eq. (1) by $N(z,M)$, integrate both sides over $\int dz \int dM$, and multiply and divide the right side by $C_{TOT}$ to obtain

$$\int dz \int dM\, N(z,M)\frac{dM}{dt} = \left[\frac{\int dz \int dM\, N(z,M)b_0(z)M^{3/4}}{\int dz \int dM\, N(z,M)M}\right] C_{TOT}$$

(2)

At steady state, *N(z,M)* is not changing in time, so we can move it inside of the derivative with respect to time, and we can also move the integrals inside of the derivative because the integration over all possible traits and masses is not a time dependent object

$$\frac{d[\int dz \int dM\, N(z,M)M]}{dt} = \frac{dC_{TOT}}{dt}$$

(3)

Furthermore, we can express the bracketed term on the right side of Eq. (1) as a mass average of a growth function as follows

$$\left[\frac{\int dz \int dM\, N(z,M)M\left[b_0(z)M^{-\frac{1}{4}}\right]}{\int dz \int dM\, N(z,M)M}\right] = \left[\frac{\int dz \int dM\, C(z,M)\left[b_0(z)M^{-\frac{1}{4}}\right]}{\int dz \int dM\, C(z,M)M}\right] = \langle b_0(z)M^{-\frac{1}{4}}\rangle_C$$

(4)

where the *C* subscript denotes that the average is taken with respect to the biomass. Combining all of this, we obtain the equation for the scaling of Net Primary Productivity or NPP equation

$$\frac{dC_{TOT}}{dt} = \langle b_0(z)M^{-\frac{1}{4}}\rangle_C\, C_{TOT}$$

(5)

This equation is in the most generic form of a TDT equation, and the growth function $f(z) = b_0(z)M^{-\frac{1}{4}}$ can be expanded such that the biomass growth equation can be expressed in terms of the biomass-weighted central moments of the trait *z*, such as the variance, skewness, and kurtosis.

We now consider a few special cases of Eq. (5) to relate to the TDT and scaling equations already in the literature. In the case that there is only a single mass value, $M^*$, or a very small range of mass values, the number density becomes $N(z,M) = N(z)\partial(M-M^*)$ in terms of a Dirac-delta function for the mass dependence. Therefore,

$\int dz \int dM\, N(z)\partial(M-M^*)M\left[b_0(z)M^{-\frac{1}{4}}\right] = \int dz\, N(z)M^*\left[b_0(z)(M^*)^{-\frac{1}{4}}\right] = (M^*)^{-\frac{1}{4}}\int dz\, C(z)b_0(z)$, and Eq. (4) becomes

$$(M^*)^{-\frac{1}{4}}\frac{\int dz\, C(z)b_0(z)}{\int dz\, C(z)} = (M^*)^{-\frac{1}{4}}\langle b_0(z)\rangle_C$$

(6)

so the scaling of NPP equation becomes

$$\frac{dC_{TOT}}{dt} = (M^*)^{-\frac{1}{4}}\langle b_0(z)\rangle_C C_{TOT}$$

(7)

The term $(M^*)^{-\frac{1}{4}}$ can be thought of as an overall normalization to the growth function $f(z)$ from TDT. As such, this result reveals that TDT, as originally formulated, essentially ignores individual mass. Thus, based on Eq. (1), growth functions within TDT should have a roughly $(M^*)^{-\frac{1}{4}}$ hidden with the normalization constant for their growth function.

Conversely, as a special case, we consider the function $b_0(z)$ to be a constant $b_0$ that occurs when z=z*. In this case, $N(z,M) = N(M)\partial(z-z^*)$ and $\int dz \int dM\, N(M)\partial(z-z^*)M\left[b_0(z)M^{-\frac{1}{4}}\right] = b_0 \int dM\, N(M)M^{\frac{3}{4}}$, and Eq. (4) becomes $b_0 \int dM\, N(M)M^{\frac{3}{4}}/C_{TOT}$. Combining these terms gives the scaling of NPP equation

$$\frac{dC_{TOT}}{dt} = \frac{b_0 \int dM\, N(M)M^{\frac{3}{4}}}{C_{TOT}}C_{TOT} = b_0 \int dM\, N(M)M^{\frac{3}{4}}$$

(8)

Following the arguments in Enquist et al. (2009), we can substitute $N(M) \propto M^{-11/8}$ to obtain $M_b \propto C_{TOT}^{8/5}$ where the subscript denotes that largest mass in the group. Using these relationships

$$\frac{dC_{TOT}}{dt} \propto b_0 \int dM\, M^{-\frac{11}{8}}M^{\frac{3}{4}} \propto b_0 \int dM\, M^{-\frac{5}{8}} \propto b_0 M_b^{\frac{3}{8}} \propto b_0 C_{TOT}^{\frac{3}{5}}$$

(9)

Defining a new constant $b_0$' to denote the product of $b_0$ with all of the proportionality constants, the overall scaling of NPP equation becomes

$$\frac{dC_{TOT}}{dt} = b_0' C_{TOT}^{\frac{3}{5}}$$

(10)

in accord with the Net Primary Productivity scaling equation derived by Enquist et al. (2009).

As a final special case, we consider the trait $z$ and the mass $M$ to be uncorrelated and the number density to be a separable function such that $N(z, M) = N(M)N(z)$. Therefore, using results from Eq. (9) and the definition of an average, Eq. (4) can be expressed as

$$\frac{\int dz\, N(z) b_0(z)}{\int dz\, N(z)} \frac{\int dM\, N(M) M^{\frac{3}{4}}}{\int dM\, N(M) M} = k \langle b_0(z) \rangle C_{TOT}^{-\frac{2}{5}}$$

(11)

where $k$ captures the proportionality constants in deriving Eq. (5). Substituting this into our overall growth equation (Eq. (5)) yields

$$\frac{dC_{TOT}}{dt} = k \langle b_0(z) \rangle C_{TOT}^{\frac{3}{5}}$$

(12)

where the average is now the standard abundance average and is *not* the biomass-weighted average. This is the growth equation in the scaling form for this special case. Alternatively, for this special case, we could express Eq. (4) as

$$\frac{\int dz\, N(z) b_0(z)}{\int dz\, N(z)} \frac{\int dM\, N(M) M \left[M^{-\frac{1}{4}}\right]}{\int dM\, N(M) M} = k \langle b_0(z) \rangle \langle M^{-\frac{1}{4}} \rangle_C$$

(13)

and the NPP scaling equation becomes

$$\frac{dC_{TOT}}{dt} = k \langle b_0(z) \rangle \langle M^{-\frac{1}{4}} \rangle_C C_{TOT}$$

(14)

This equation is complete equivalent to Eq. (13) but expresses the growth function more in terms of the TDT framework such that the right side appears to have an overall linear dependence in $C_{TOT}$, and as a result, we have a mixture of types of averages, with the function $b_0(z)$ being abundance averaged and the $M^{-1/4}$ being biomass averaged.

The major results of this section are Eq. (5), which is the most general formulation of the growth equation because it does not rely on traits or mass being constant or uncorrelated. In this form, the growth equation is like the TDT formulation, but as the special cases below it reveal, $\langle b_0(z) M^{-\frac{1}{4}} \rangle_C$ may hide extra dependencies on $C_{TOT}$. Eq. (7) is the result when the mass is constant and is expressed in the form of TDT equations such that it is linear in $C_{TOT}$ and reveals an overall $M^{-1/4}$ for adjusting the growth function across groups. Eq. (10) is the special case where the traits are constant and reduces to the exact scaling equation given in Enquist et al.(2009). Finally, when traits and mass are uncorrelated, Eq. (13) and Eq. (14) are two different but completely equivalent ways to express the growth function. Eq. (13) is in the form of scaling equations by consolidating the mass average with the $C_{TOT}$ dependence, while Eq. (14) is in the form of TDT equations by keeping two averages around, including one that is an abundance average and one that is a biomass-weighted average. For all of these equations and cases, the functions inside the averages can be expended in terms of moments as done for TDT for biomass-weighted averages or as done in (2004), for abundance-weighted averages.

## III. Growth functions across environmental gradients: Incorporating trade-offs into TDT

Importantly, as discussed in the main text, equation 9 predicts an unbounded growth response such that increasingly larger values of $b_0$ always leads to increased growth, which realistically cannot continue indefinitely. A key assumption of TDT is that there is fundamental tradeoff between a given trait value and the performance of an organism across an environmental gradient, $E$. The final step to integrate a general TDT that can link traits, organismal performance, and environmental gradients, is to specify tradeoffs between underlying traits, growth, and metabolic scaling.

Within a given environment, $E$, an important question is what would prevent $\bar{b}_0$ from becoming infinitely big? or small? In the case of growth rate, possible tradeoffs likely include the types of limiting resources individuals use or the environmental conditions for optimal growth. So, individuals that allocate internal resources to specific traits defined by $b_0$ may reduce the impact of one limiting environmental factor but this would necessarily incur a disadvantage with respect to another environmental limiting factor.

A trade off or cost function can be formulated within the growth function, $f$. Multiplying this cost function by $f$ shows that, for a given $E$, the growth function has a maximum at $z_{opt}$ or here $b_{0_{opt}}$ and, as a result, the second derivative of $f$ (the second term in eqn. 5 and 14), will be negative, as long as $\bar{b}_0$ is close to $b_{0_{opt}}$.

We can specify a generic form of a trade off by following Norberg et al. (2001). We can approximate a trade off by first invoking a general quadratic or Gaussian cost function on the community value ($\bar{b}_0 - b_{0_{opt}}$). We add a cost function to eqn. 5 and 14. This provides a general form of a trade off. That new cost function that is multiplied by $f$ could be a general quadratic $\left[1 - \left(\frac{\bar{b}_0 - b_{0_{opt}}}{\sigma^2}\right)^2\right]$ or Gaussian function, $\exp\left[-\left(\frac{\bar{b}_0 - b_{0_{opt}}}{\sigma^2}\right)^2\right]$. Here, $\sigma^2$ is the observed standard deviation in the trait or $b_0$ observed within the assemblage. Both cost functions, reduce to 1 when $\bar{b}_0 = b_{0_{opt}}$ (e.g. for a given environment $E$, the observed mean community trait, $z_{opt}$, and average metabolic normalization, $\bar{b}_0$ is at the local optimum). Note, both decrease in value as you go away from $b_0 - b_{0_{opt}}$. Dividing through by $\sigma^2$ defines the penalty for individual growth rate for being away from the optimum. Thus, for a given environment, $E$, characterized by a unique $b_{0_{opt}}$, the growth function can be made more explicit in terms of a generic trade-off where

$$f(b_0) = c_0 b_0 \langle M^{\theta-1}(b_0) \rangle \left(1 - \frac{(b_0 - b_{0,opt})^2}{\sigma_{b_0}^2}\right)$$

(15)

This is a modified growth function and is characterized in Fig. 2. It can be made more specific by incorporating the traits that then define $b_0$. Here, the first term is the general growth equation from the relative growth rate literature (Poorter, 1989) that has been more formally derived in metabolic scaling theory. The second term of eqn S1 is the associated tradeoff. As a result, the second term in the TDT eqn 5 in the main text gives how much of whole-community biomass production is *reduced* due to the amount of trait variance, $V$, in the community because of the explicit trade off function, the second derivative of $f$, $\frac{d^2 f_{b_0=\bar{b}_0}}{db_0^2}$, would then be negative near $b_{0_{opt}}$. As a result, increasing variance in $b_0$ would then *decrease* total community production.

### IV. Methods: Approximating the shape of the community trait distribution via community weighted measures.

In order to assess predictions of Trait Drivers Theory, TDT, it is necessary to quantify the full distribution of traits in a community, $C(z_i)$. This involves measuring the trait values of *all individuals* and thus incorporates both inter- and intraspecific trait variability. While measuring traits of *all* individuals in a community is ideal and several studies have done so (Gaucherand & Lavorel, 2007; Lavorel *et al.*, 2008; Albert *et al.*, 2010), it is time consuming work (Baraloto *et al.*). While there are limitations, the trait biomass distribution, $C(z_i)$, can be approximated and predictions of TDT can be tested without explicitly measuring the traits of all individuals.

Trait distributions can be approximated in two ways. The first method is straightforward and calculates the weighted trait distribution by taking the mean species trait value and multiplying by a measure of dominance (cover, biomass, abundance (Grime, 1998)). This method can be implemented by calculating the central moments of joint-distribution.

This community weighted variance or CWM is increasingly a standard metric in trait-based ecology (Violle *et al.*, 2007; Garnier & Navas, 2012; Lavorel, 2013) and represents the trait mean calculated for all species in a community weighted by species abundances as follows:

$$CWM_{j,y} = \sum_{k=1}^{n_j} A_{k,j} \cdot z_k$$

(16)

where $n_j$ is the number of species sampled in plot $j$, $A_{k,j}$ js the relative abundance of species $k$ in plot $j$, and $z_k$ is the mean value of species $k$. Several studies have also assessed the community weighted variance of the trait distriubiton (see (Lavorel *et al.*, 2011; Ricotta & Moretti, 2011).

The assemblage variance, $V$ is calculated via the biomass-weighted values for the community weighted varince ($CWV_{j,y}$) is given by,

$$CWV_{j,y} = \sum_{k=1}^{n_j} A_{k,j} \cdot \left(z_k - CWM_{j,y}\right)^2$$

(17)

Further, the central moments skewness and kurtosis ($CWS_{j,y}$ and $CWK_{j,y}$, respectively) are given by

$$CWS_{j,y} = \frac{\sum_{k=1}^{n_j} A_{k,j} \cdot \left(z_k - CWM_{j,y}\right)^3}{CWV_{j,y}^{3/2}} \; ; \quad CWK_{j,y} = \frac{\sum_{k=1}^{n_j} A_{k,j} \cdot \left(z_k - CWM_{j,y}\right)^4}{CWV_{j,y}^2} - 3$$

(18)

A limitation of this approach, however, is that it ignores the contribution of intraspecific trait variability. Community trait moments may also be sensitive to the distribution of abundances across species. For example, a highly positive community kurtosis value may just reflect the hyper-dominance of one species and not the true dispersion of traits again due to intraspecific variation.

A second method utilizes sub-sampling individuals to obtain a better approximation of how intraspecific variation influences the community distribution. By subsampling individuals for each species one can begin to incorporate intraspecific variation around mean trait values for each species.

In Figure S5 we highlight a typical example that we believe can be used to generate two approximations of the community trait distribution. We use data from Konza Prairie LTER, Kansas, USA (McAllister *et al.*, 1998). First, data were collected for the abundance of each species. These data are illustrated in the Figure S4 to estimate the community trait distribution from mean and variance measure of species traits. We find that, consequently and counter-intuitively, the inclusion of intra-specific variation will likely simplify modeling efforts because these types of distributions are much easier to manipulate and understand analytically. For each species, the standard deviation of trait variation is equal to the reported standard error multiplied by the square root (where *n*=3). In sum, the community trait distribution can be approximated in two ways (methods B and C). While **B** emphasizes interspecific variation, **C** also begins to include intraspecific variation. Method **B** is a reasonable approximation and can easily be implemented by most ecological studies as it only requires interspecific trait information and local abundance values. Method **C** requires an additional standardized sub-sampling method to estimate the standard error for each species but will result in a more accurate moment approximation.

**V. Methods: Rocky Mountain Biological Lab: Shifts in trait distributions and ecosystem measures across an elevational gradient –**

*Measuring whole-community trait distributions* - We measured community trait distributions and whole- ecosystem carbon flux data along an elevational gradient near

Gothic, Colorado. The elevation gradient ranged between 2,460 to 3,380 m and spans 39 km in geographic distance. The elevational gradient contains five long-term study sites that run from dry, shrub-dominated high desert in Almont Colorado (2475 m) through the subalpine zone, to just below tree line (3380 m). The gradient consists of five long-term study sites that were established by B.J. Enquist in 2003 and has been sampled every year since. The gradient is located within Washington Gulch and East River valleys near the Rocky Mountain Biological Lab.

Each study site along the elevation gradient has similar local slope, aspect, and vegetation physiognomy. The sample area is approximately 50 m$^2$ in area and consist of a mixture of shrubs, grasses, and forbs. As discussed in Bryant et al. 2008 (Bryant *et al.* 2008) there is substantial turnover of plant species between sites with very few of the 120 species sampled occurring in more than two of the sites. Additionally, shrub cover across the gradient decreases from a high of 33% at the lowest elevation site to 0% at the highest.

We utilized carbon flux data collected during the summer months of 2010 measured across the gradient. A species list and phylogeny for species at each site is given by Bryant and others (2008). All sites contain weather stations on-site or nearby. Each study site has a similar local slope and south-southwest aspect, and contains a mixture of herbaceous perennials, grasses, and shrubs. Since 2003, each year, five 1.3 m x 1.3 m plots have been established haphazardly along the local slope of each study site, with at least five meters distance between plots.

In 2010 Henderson et al. measured the SLA of in each plot and collected one fully expanded leaf from every individual. Fresh leaf samples scanned (with petiole) in the laboratory then dried to a constant mass and weighed. The trait values measured from every individual in each community were compiled to create individual level trait distributions for SLA. In total, leaves from 2,253 individuals across 54 species were collected and measured at the five sites. Species turnover was high, with only 11 species being found at more than one site and only one species found at more than two sites.

*Gas exchange and productivity measures –* In 25 plots (5 plots per elevational site) we measured total ecosystem carbon flux. Carbon flux was measured as instantaneous daytime peak uptake (ca. 10 am) and nighttime peak respiration (ca. 10 pm) (Saleska and others 1999). Ambient $CO_2$ was measured by a Li-Cor 7500 infra red gas analyzer for 30 seconds, and then the tent was put in place over the plot and the $CO_2$ concentration within the tent was measured for 90 seconds (Jasoni and others 2005). Daytime measurements were only taken under cloudless conditions. The tent was designed to let in 75% of photosynthetically active radiation (tent fabric by Shelter Systems). Air inside the tent is well mixed by fans, and the tent chamber was sealed using a long skirt along the base of the tent that that was covered with a heavy chain. The volume of the tent used along the gradient was 2.197 m$^3$.

Soil efflux was measured at the same time as NEP using a Li-Cor 6400 portable photosynthesis machine with the soil chamber. The soil chamber fit inside a PVC soil collar, which was placed in the plot at least two weeks prior to the first measurement. Soil efflux was measured in two places in each plot along the gradient and one place per plot for the manipulation.

Carbon flux measurements along the elevational gradient were taken four weeks after snowmelt and then again at peak season (approximately four weeks after the first measurement, or when the majority of plants reached maximum height). Each NEP measurement consisted of daytime peak uptake (at ~10:00) and nighttime respiration (at 22:00). Following the method of Jasoni *et al.*, (2005) ambient $CO_2$ was measured for 30 seconds, and then the tent was lowered and the $CO_2$ concentration within the tent was measured for 90s, under clear sky and low wind conditions. Air inside the tent was well mixed by fans, and the chamber was sealed to the ground using a heavy chain.

For data analysis we fit the predicted Trait Drivers Theory model, using multiple regression in R using the 'car' library we fit the following linear model

```
lm(log10(NEP) ~ CWM.SLA + CWV.SLA + logBio + as.factor(Site)
```

where CWM.SLA is the community weighted mean SLA and CWV.SLA is the community weighted variance SLA calculated using the above equations for CWM and CWV as presented in the vegan package in R. Here, all values of SLA were log transformed before analyses and logBio is the log10 total above ground biomass at time of carbon flux measurement. We use the site elevation of the sample as a factor in the model. The fit of this model $r^2$=0.778, df=22, F=11.04, $p$ < 0.0001, AIC -24.38. All variance inflation factors were generally less than 5 except for Site where the vif = 5.86 which is still low for a vif value. In this model the effect of CWV.SLA is significant (*p*= 0.023, *t* = 2.45, SE= 0.072 parameter 0.337) but CWVSLA and total biomass is marginally significant (*p*= 0.068, t = -1.921, se = 0.067, parameter= -0.138; p=0.059; *t* = 1.994; se= 0.067, parameter = 0.135).

Variation in Net Ecosystem Production across the gradient appears to be primarily due to the CWM and CWV of community SLA. Removing the parameter biomass and fitting a more simplified model with only mean and variance in SLA,

```
lm(log10(NEP) ~ CWM.SLA + CWV.SLA + as.factor(Site)
```

predicts a similarly amount of variation in NEP to the full model above (p < 0.0001, $r^2$= 0.7384 , AIC =-21.40 and both CWMSLA and CWVSLA are now significant within the model (*p*= 0.035, *t*= 2.247, se= 0.358, parameter = 0.803; p= 0.0386, t= -2.194, se = 0.075, parameter = -0.165)).

Fitting a more simple model just using either plot biomass or plot CWM.SLA with site as a factor results a poorer fit model when compared with the TDT predicted model with lower $r^2$ and higher AIC values ( (log10(posNEP)~ log10Bio +as.factor(Site_name, $r^2$= 0.687 , AIC=17.994; lm(log10(posNEP)~ CWM.SLA +as.factor(Site_name); $r^2$= 0.684, AIC = -17.704). Further, in both models, the effect of biomass and CWMSLA were marginally significant (*p*= 0.064 and *p*= 0.056  respectively). These results indicate that together the CWM and CWV of community SLA are primary drivers of variation in community carbon flux.

## VI. Methods: Park Grass Experiment: Background, Methods, & Discussion

*Background and Methods* - The original purpose of Park Grass Experiment (PGE), started in 1856, was to investigate the effects of high levels of inorganic fertilisers and organic manure on hay production relative to control treatments (see references within Crawley *et al.*, 2005) for additional details on methodology). Our analyses mainly focused on the PGE trait dynamics of Plots 2 and 16. These plots were selected because of their contrasting botanical composition and species richness (Crawley *et al.*, 2005; Harpole & Tilman, 2007). Plot 2 (became plot 2/2 in 1996) received farm yard manure between 1856 and 1863, but since then has received no further manure or fertiliser inputs, and is now considered to be a control plot. Plot 16, started in 1858, is a fertilized, unlimed plot that receives annual N, P, K, Na, Mg applications (48kg N ha$^{-1}$/ as sodium nitrate in spring; mineral applied in winter: 35kg P ha$^{-1}$ P as triple superphosphate, 225 kg K ha$^{-1}$/as potassium sulphate, 15 kg Na ha$^{-1}$ as sodium sulphate, 10 kg Mgha$^{-1}$as magnesium sulphate). For the Park Grass dataset we approximated the central-moments of the community trait distribution, $C(z)$ for trait $z$ within plot $j$ and year $y$ using equations 9-11. We analyzed the time-series of these plots in terms of changes in botanical composition, traits, and species richness. To focus on how experimentally paired *local* communities have responded over time we highlighted plots 2 and 16 (the other plots also showed similar responses).

*Assignment of trait values and biomass weighted trait distributions* - We assessed changes in specific leaf area (SLA), seed size, and height. These traits have also been proposed to capture most functional and life history variation across species (Westoby, 1998; Westoby *et al.*, 2002). Seed size is thought to characterize regenerative traits not associated with our trait-based growth model developed in eqn 7. Including a regeneration trait provides a basis to assess if other niche or dispersal based processes acting on other traits may be more important in structuring the community than traits associated with growth, $dC/dt$ (see also discussion on effect and response traits (Suding *et al.*, 2008)). Further, variation in seed size should not directly influence our ecosystem level predictions for $dC_{Tot}/dt$ as this trait is not explicit in eqn 7. According to eqn. 7, plant height (or size, $C$) can influence ecosystem NPP. So any large shifts in mean plant height would be important to note as well. Trait values are for populations sampled in UK. We used the first four central moments of $C(z)$ for plot $j$ and year $y$, to calculate the biomass-weighted mean, variance, skewness, and kurtosis.

Within this experiment, species abundance was measured by cutting aboveground biomass to ground level from six randomly located quadrats (50cm x 25cm) within each experimental and control plot. The plant material was then sorted into species, oven dried at 80 degrees C for 24 hours, and the dry matter determined (Williams, 1978; Crawley *et al.*, 2005). For each plot, yields were estimated by weighing standing biomass (t/ha at 100% dry matter) from the whole plot, harvested in mid-June. The plots were originally cut by scythe, then by horse-drawn, and then tractor-drawn mowers (Williams, 1978; see these references for additional methdlogical detail of the PGE); Crawley *et al.*, 2005).

Our analysis of additional fertilization and control plots at Park Grass (plots 3 & 14) also reveal similar differences in trait distributions (mean, variance, skew, and kurtosis) between the control and fertilized plots. In sum, for all of the Park Grass plots, our central conclusions do not change. We observed coordinated shifts in the functional trait distribution. In Table S1 we show the correlations between the central moments of the trait distribution and species richness. These correlations include plots 2 (control) & 16 (fertilized) together. *None* of the central moments have significant correlation with species richness, indicating that the mechanisms and responses to environmental change captured by the shape of trait distributions are not captured by species richness. Figure S1 shows the change in the central moments of the community trait distribution for SLA or the specific leaf area. Figure S4 shows the associated changes in seed size in the 140-year long-term Park Grass experiment. Figure S5 shows the change in the central moments of the community trait distribution for adult height in the 140-year long-term Park Grass experiment

*Park Grass Experiment or PGE – Additional Discussion -* Fertilization also changed the shape of the specific leaf area (SLA) trait distribution indicating that the underlying forces that structure these communities under differing environments changed. For example, fertilization led to a reduction in the variance (the community mean SLA was negatively related to the community SLA variance, $r^2= 0.48$, $P<0.001$), suggesting either that fertilization was an environmental 'filter' on traits and/or competitive exclusion increased (see Table 2). The observed increase in the skewness and kurtosis of the SLA distribution with fertilization is in accord with predictions and expectations of TDT where more quick directional shifts in $z_{opt}$ will lead to a skewed distribution. In contrast, the control plot trait distribution did not show dramatic changes in the variance or skewness of the distribution. However, the mean of SLA in the control plot did significantly decrease suggesting that natural and / or more gradual changes in the local environment (possibly due to an increase in nitrogen limitation over time and/or climate change) and/or recovery from past disturbance has influenced the control community. The kurtosis of the trait distribution in the control plot remained negative and close to -1.2 (a uniform distribution) consistent with an increased role of divergent ecological forces (niche packing and the role of biotic interactions). In contrast, in the fertilized plot, the variance of the distribution has decreased and the kurtosis tends to exhibit high positive values.

In the PGE fertilization can be seen as a specific environmental driver, *E*. Fertilization changes soil resource availability and, according to TDT, differentially shifts the optimum growth rate. Indeed, in support we see a shift in $z_{opt}$ (here being SLA) associated with fertilization. Analysis of the moments of distributions for two other community traits, seed size and adult reproductive height, show that these trait means did not appreciably change with fertilization (Fig S4-S5). Importantly, the community mean of plant height did not change with fertilization supporting our assumption that observed change in community NPP was primarily due to changes in SLA, and also that the mean plant size or biomass, $C$, did not appreciably change. The one change with fertilization that we do observe is that the community variance of plant height increased. No other traits show any changes in the fertilized plots. Future work elaborating TDT should include the role of multiple trait drivers and their associated predictions.

**Table S1.** Correlations between the central moments of the community trait distribution of specific leaf area or SLA and species richness, (Plots 2 (control) & 16 (fertilized) together). None of the moments have significant correlation with species richness, indicating that the mechanisms and responses to environmental change captured by the moments are not captured by species richness. This represents one of the great advantages for Trait Driver Theory (TDT) over theories based on species richness.

|  |  | Species richness |
|---|---|---|
| SLA | mean | $0.00^{ns.}$ |
| | variance | $0.05^{ns.}$ |
| | skewness | $0.08^{ns.}$ |
| | kurtosis | $0.03^{ns.}$ |

**Figure S1** – Trait Driver Theory and examples of the first four central moments of trait distributions: **A)** *mean* (the first moment) and *variance* (the second moment). In this example, we show two communities with the same mean trait value (dashed line) but with different variances; **B)** *skew* (a combination of the second and third moments). Skewness in a trait distribution can be caused by (i) time lags in community response to a new optimum trait value where a long-tail of individuals expressing suboptimal trait values is present in the community (e.g. see Fig 2), (ii) by lopsided trait immigration into the community, or (iii) by physical or physiological limits on trait expression (e.g. hydrological constraints on plant height); and (iv) may reflect rare species advantage. As shown in this example; and *Kurtosis* (a measure of the fourth moment relative to the second moment). Competitive exclusion and/or strong stabilizing selection will give a highly peaked ($4^{th}$ moment kurtosis) distribution (**C**) while niche packing reflecting biotic interactions could give a more uniform distribution (**D**). Note, a normal trait distribution is defined by a skewness and kurtosis = 0. The more peaked the distribution the more positive the kurtosis value (including the logistic, hyperbolic secant, and Laplace distributions). In contrast, processes that result in the 'spreading out' of traits will be characterized by increasingly more negative kurtosis values. In the case of a uniform distribution, kurtosis = -1.2. An increasingly bimodal distribution (Bernoulli distribution) will have kurtosis values = -2. (**E**). Bimodal distributions could arise where there are multiple optimal (dashed lines) trait values (**F**), or where the community is responding to a recent environmental change where the two peaks represent both an increased representation of nearly optimal individuals (the high peak) and the continued presence of individuals with optimal trait values for the historic environment (the low peak) (**G**). Dashed lines correspond to the optimal trait value(s).

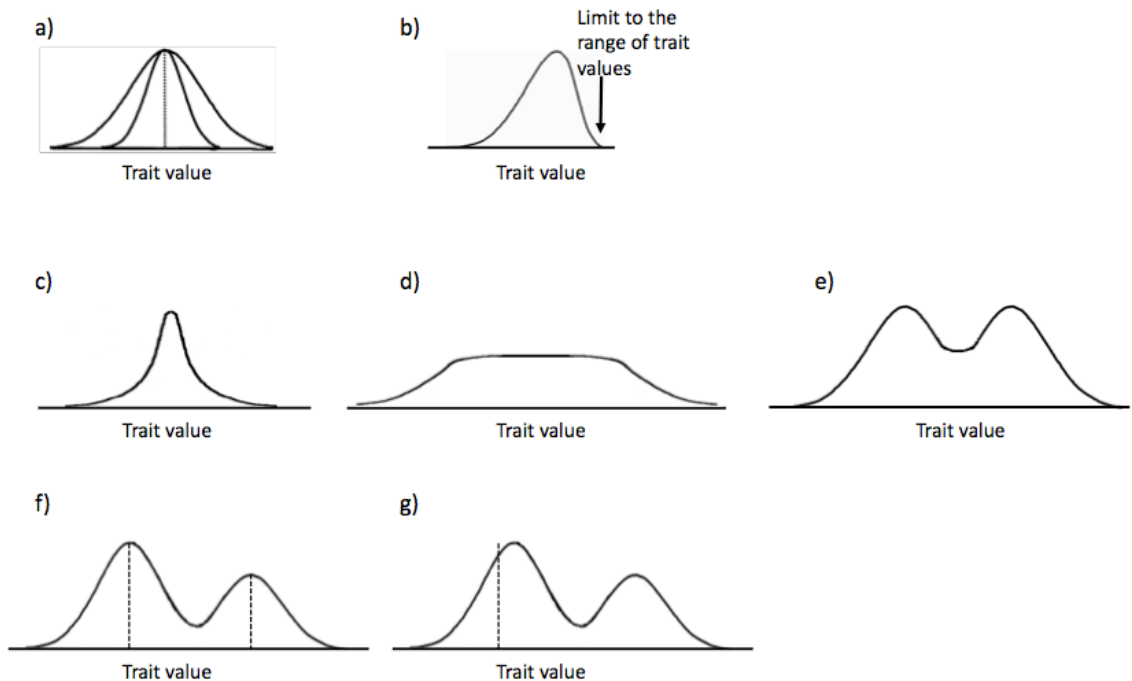

**Figure S2**. Examples of estimation of a community trait distribution from utilizing either mean trait values and/or intraspecific variation. In this example we use abundance information (**A**) for the percent cover for 16 species from Konza Prairie LTER, Kansas, USA. Next, a trait measure, the leaf photosynthetic rate, was measured on a minimum of three leaves on three separate plants. Species means and standard errors were then calculated for this trait for each species. In (**B**) using the first method, the trait abundance distribution was calculated using only the mean trait data for the species in **A**. Numbers in the parentheses indicate the number of species in each trait bin, and the peaks correspond to some of the dominant species in plot (A) that were rank ordered by abundance and not by photosynthetic rate. Lastly, using the second method, the community trait distribution of all individuals can be further approximated by integrating intraspecific subsampling. In (**C**), for each species, we incorporated intraspecific variation by using the standard error for this trait as measured in each species and then assumed a normal distribution around each species mean. We then generated the community-wide trait distribution by sampling from each species intraspecific trait distribution (defined by its mean and SE). The resulting distribution (C) is much more continuous and unimodal than in plot (B), which does not include intra-specific variation.

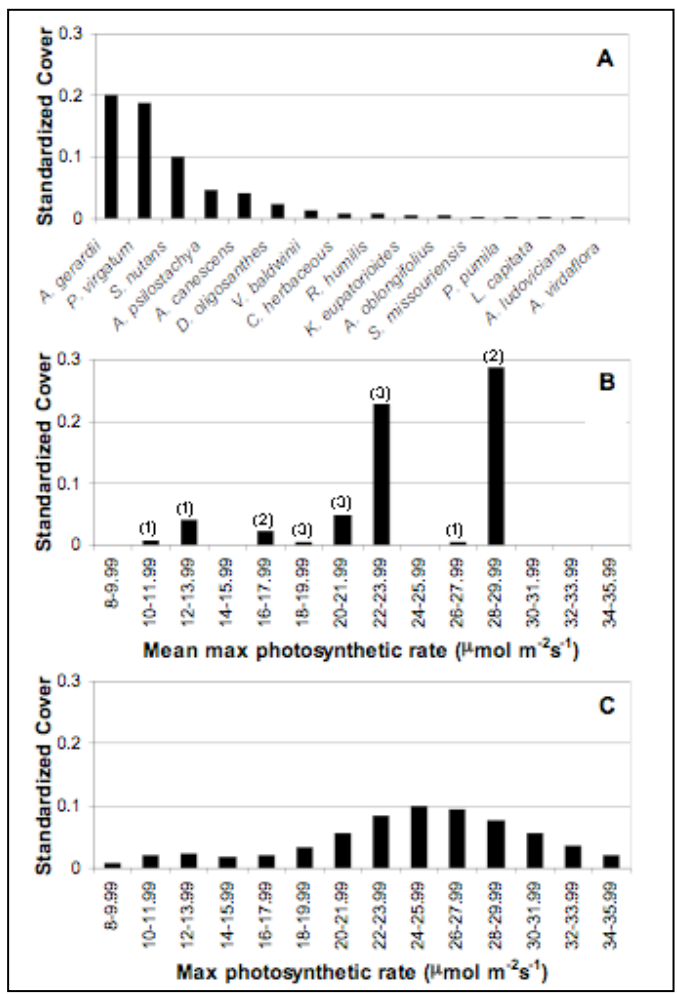

**Figure S3.** Change in the central moments of the community trait distribution for a key trait – specific leaf area or SLA - in the 140-year long-term Park Grass experiment. Regression lines are indicated for significant relationships. Fertilization has caused a decrease in the variance and an increase in the skewness. Fertilization increases the mean assemblage specific leaf area (SLA) but reduces the variance. This result indicates a directional shift in trait optimum, $z_{opt}$, and a functional shift in the composition of the community. Further, fertilized plots have become increasingly more skewed and have increasingly more positive kurtosis values indicating that communities have become increasingly dominated by a few trait values. Distributions with kurtosis values of -1.2 are characteristic of an 'overdispersed' uniform distribution while plots with kurtosis values greater than 0 are more clumped/peaked than expected from a normal or Gaussian distribution.

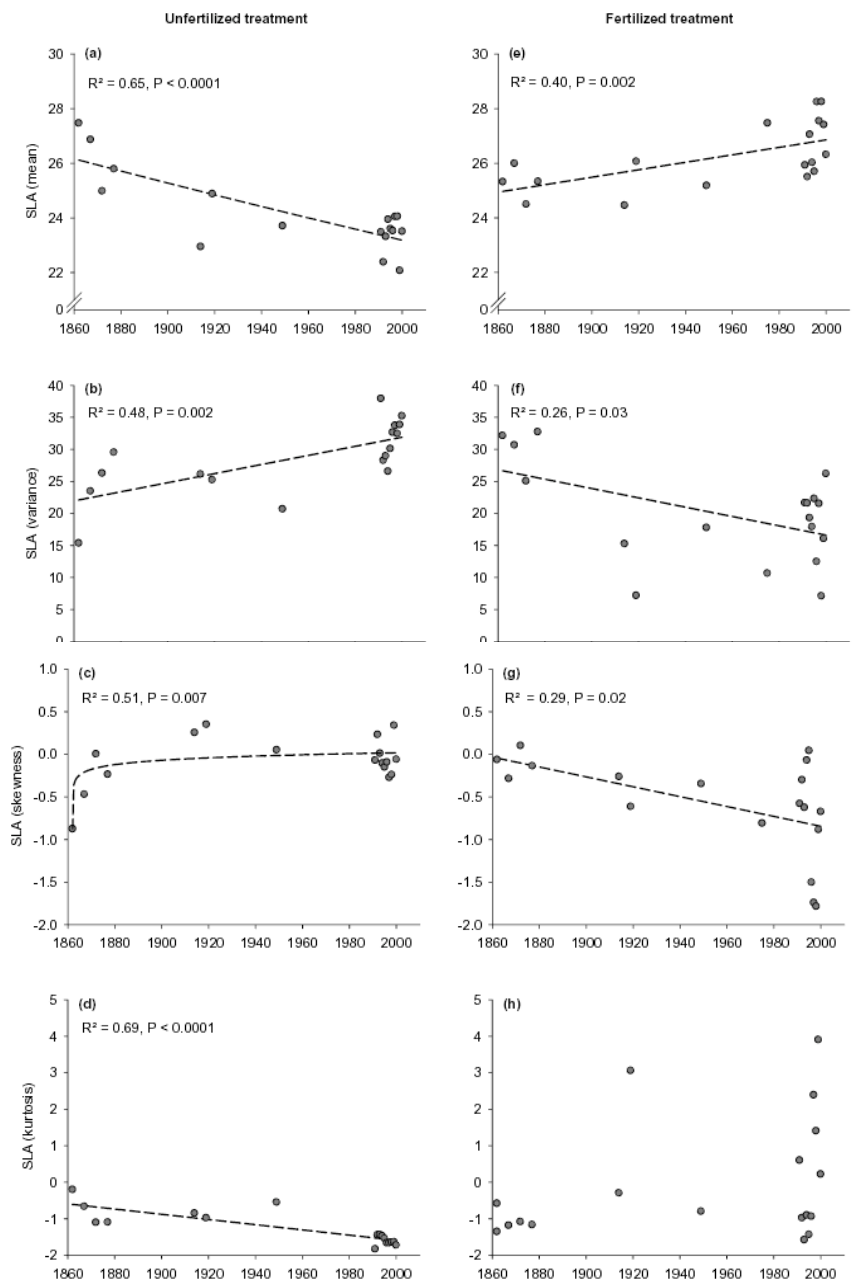

**Figure S4**. Change in the central moments of the community trait distribution for seed size in the 140-year long-term Park Grass experiment. Significant correlations are indicated with presence of fitted (dashed) regression lines.

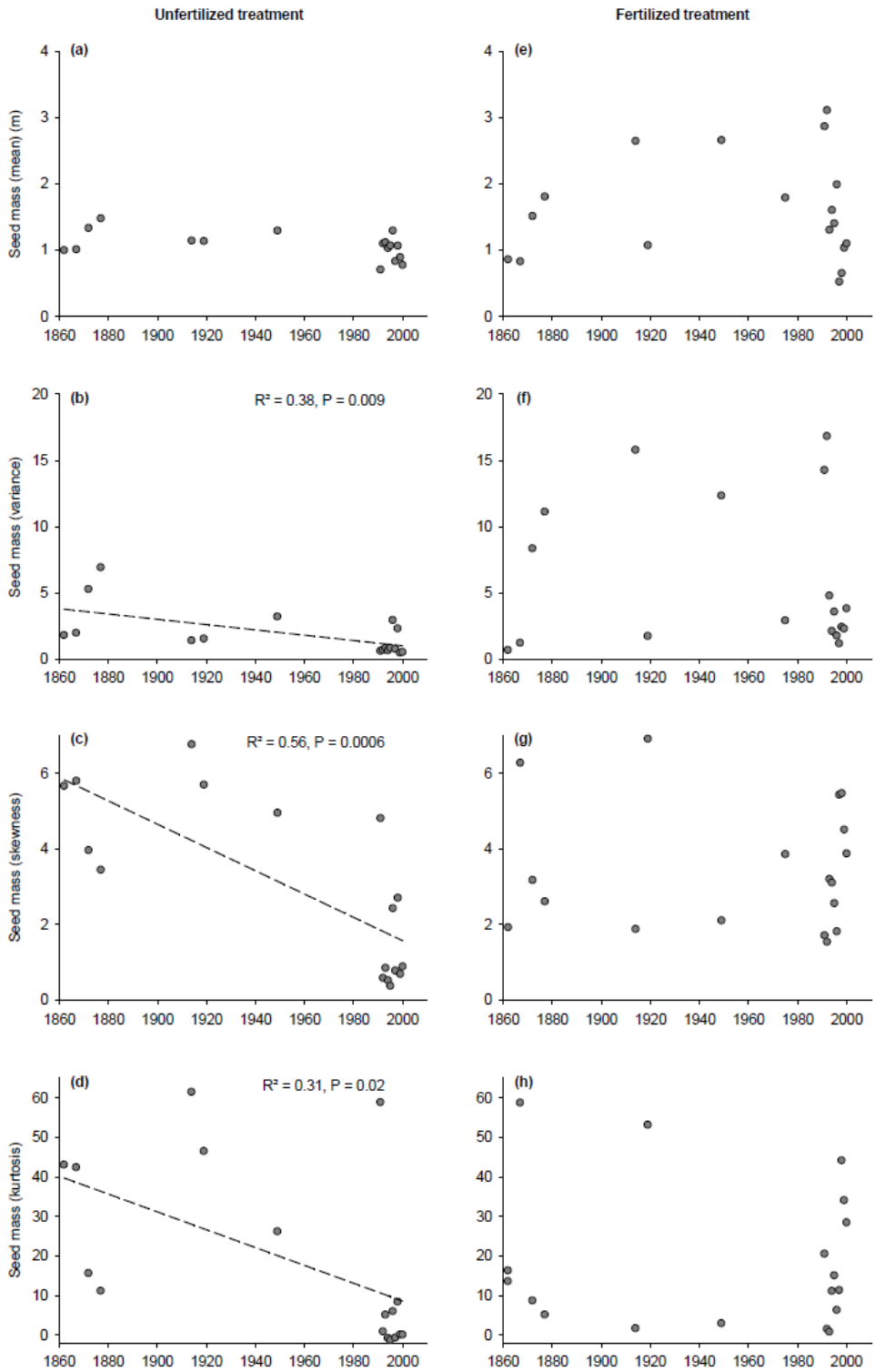

**Figure S5**. Change in the central moments of the community trait distribution for adult height in the 140-year long-term Park Grass experiment. Significant correlations are indicated with presence of fitted (dashed) regression lines.

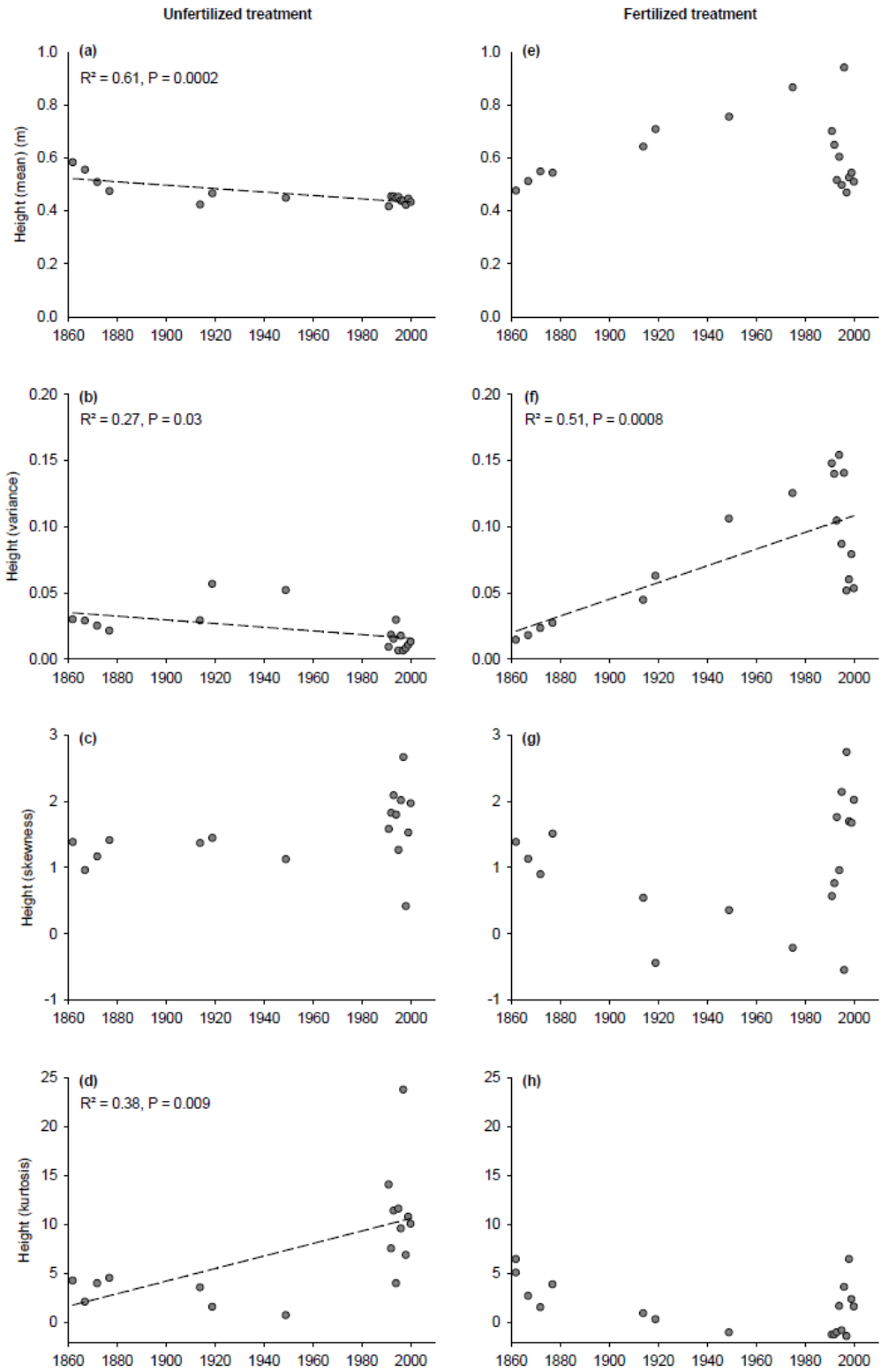